\documentclass[prb,showpacs,twocolumn,aps,superscriptaddress,floatfix]{revtex4-1}
\usepackage{amsmath}
\usepackage{amssymb}
\usepackage{bm}
\usepackage{graphicx}
\usepackage{color}
\usepackage[svgnames]{xcolor}
\usepackage{soul}
\usepackage{hyperref}
\usepackage{verbatim}

\graphicspath{{FIGS/}} 

\begin{document}

\title{Triplet superconductivity and spin density wave in biased AB bilayer
graphene}

\author{A.O. Sboychakov}
\affiliation{Institute for Theoretical and Applied Electrodynamics, Russian
Academy of Sciences, 125412 Moscow, Russia}

\author{A.V. Rozhkov}
\affiliation{Institute for Theoretical and Applied Electrodynamics, Russian
Academy of Sciences, 125412 Moscow, Russia}

\author{A.L. Rakhmanov}
\affiliation{Institute for Theoretical and Applied Electrodynamics, Russian
Academy of Sciences, 125412 Moscow, Russia}

\begin{abstract}
We examine spin density wave and triplet superconductivity as possible
ground states of the Bernal bilayer graphene. The spin density wave is
stable for the unbiased and undoped bilayer. Both the doping and the applied
bias voltage destroy this phase. We show that, when biased and slightly
doped, bilayer can host a triplet superconducting phase. The mechanisms for
both ordered phases rely on the renormalized Coulomb interaction.
Consistency of our theoretical conclusions with recent experimental results
are discussed.
\end{abstract}


\date{\today}
\maketitle

\section{Introduction}

Experimental observation of Mott insulating states and superconductivity in the
magic-angle twisted bilayer
graphene~\cite{NatureMott2018,NatureSC2018,MottSCNature2019}
encouraged further studies of correlated phases in
bilayer~\cite{bilayer_review2016}
and multi-layer graphene systems. The most well-researched type of bilayer
graphene is AB, or Bernal, bilayer graphene (AB-BLG). There is experimental
evidence~\cite{GapExp1PNAS2012,GapExp2NatureNano2012,GapExp3PRB2012,
GapExp4PRB2013}
that the ground state of AB-BLG is gapped even at zero bias voltage and
zero doping, and the gap is of many-body nature. The kind of the ground
state hosted by AB-BLG is under discussion. Different candidates for this
low-temperature phase, such as
ferromagnetic~\cite{FerromagneticBLG2006},
spin-density wave
(SDW)~\cite{antiferromagnetic1BLG2012,antiferromagnetic2BLG2013,
rakhmanov2023QM},
``pseudo magnetic''~\cite{PseudoSpinMagneticBLG2008},
nematic~\cite{NematicPRB2010},
among other possibilities, have been proposed.

Recently, a cascade of transitions between several non-superconducting
states~\cite{CascadeNature2022,SCBLGNature2022,Seiler2022},
as well as the
superconductivity~\cite{SCBLGNature2022},
have been observed in the doped and biased AB-BLG. In
Ref.~\onlinecite{rakhmanov2023QM}
we argued theoretically that the transitions cascade reported in
Refs~\onlinecite{CascadeNature2022,SCBLGNature2022,Seiler2022}
can be connected to the sequence of several fractional metallic states (with
spin and valley polarizations) that become stable in the doped SDW phase.

As for the AB-BLG superconducting phase, its transition temperature was
experimentally estimated to be
$T_c\approx26$\,mK.
Curiously, the superconductivity appeared only when the magnetic field of
about $150$\,mT is applied parallel to the bilayer. To explain the
superconductivity in AB-BLG, both
phonon~\cite{PhononSC1DasSarma2022,PhononSC2DasSarma2022}
and electronic
mechanisms~\cite{ElectronSCRoyPRB2022,Guinea2023superconductivity,
wagner2023superconductivity,LevitovSCPRB2023,ElectronSCCeaPRB2023,
LevitovSCPRB2023}
have been proposed.

Unlike our previous
paper~\cite{rakhmanov2023QM},
which was dedicated to the non-superconducting states of the AB-BLG, here
we focus on the superconductivity in the same system. Our starting point is
the usual four band tight-binding model with Coulomb
interaction~\cite{bilayer_review2016}.
The model is studied using zero-temperature mean-field approximation. To
account for screening, the renormalized Coulomb potential is calculated
within the random phase approximation (RPA). In contrast to similar
approaches (see, e.g.,
Refs.~\onlinecite{Guinea2023superconductivity,
wagner2023superconductivity}),
we use the tight-binding model and distinguish intra-layer and inter-layer
Coulomb potentials which, as demonstrated below, experience dissimilar
screening.

Our analysis is started with the mean-field study of the SDW phase in the
undoped unbiased bilayer. Typically one expects that the SDW phase is more
robust than a superconductivity, which is indeed consistent with our
findings. Thus, the SDW must be weakened to allow for the stabilization of
the superconductivity. Application of the bias voltage and doping favor the
superconductivity. We prove that the renormalized Coulomb potential is
enough to stabilize the triplet superconducting $p$-wave pairing in the
AB-BLG. Our estimates for the superconducting state properties, and in
particular
$T_c$,
are consistent with the experiment.

The paper is organized as follows. In
Sec.~\ref{sec::TB_model}
the tight-binding Hamiltonian is described. Renormalized Coulomb
interaction in the unbiased undoped AB-BLG is calculated in
Sec.~\ref{sec::undoped_RPA}.
We study the SDW phase in
Sec.~\ref{sec::MF_SDW}.
Renormalized interaction for the doped biased bilayer is calculated in
Sec.~\ref{sec::doped_RPA}.
Section~\ref{sec::superconducting_mech}
is dedicated to the superconducting phase. More informal discussion of our
findings, as well as conclusions of our analysis, can be found in
Sec.~\ref{sec::discussion}.
Specific technical details are placed in two appendices.

\section{Tight-binding model}
\label{sec::TB_model}

In the AB-BLG, carbon atoms in sublattice $B$ of the top layer are located
right above the atoms of the sublattice $A$ of the lower layer, while the
atoms in sublattice $A$ of the top layer are located above centers of
hexagons formed by the atoms of the lower layer. There are four atoms
per unit cell. The elementary translation vectors for the AB-BLG can be chosen as
$\mathbf{a}_{1,2}=a(\sqrt{3},\mp1)/2$,
where
$a=2.46$\AA\,
is the elementary unit length. Vector
$\bm{\delta}=(\mathbf{a}_1+\mathbf{a}_2)/3$
connects two atoms within a single unit cell in the same layer. The
inter-layer distance for AB-BLG is
$d=3.35$\,\AA.

We consider the following model Hamiltonian
$H=H_0+H_{\text{int}}$,
the first term being the single-particle Hamiltonian, while the second term
describing the Coulomb interaction. These are
\begin{eqnarray}
\label{Hkin}
H_0
&=&
\sum_{\mathbf{k}\sigma}
	\psi^{\dag}_{\mathbf{k}\sigma}
	\left({\cal H}_{\mathbf{k}}-\mu\right)
	\psi^{\phantom{\dag}}_{\mathbf{k}\sigma},
\quad
\\
\label{Hint}
H_{\text{int}}
&=&
\frac{1}{2\cal N}\!
\sum_{\mathbf{kk}'\mathbf{q}\sigma\sigma' \atop ij\alpha\beta}
	d^{\dag}_{\mathbf{k}+\mathbf{q}i\alpha\sigma}
	d^{\phantom{\dag}}_{\mathbf{k}i\alpha\sigma}
	V^{ij}_{\mathbf{q}}
	d^{\dag}_{\mathbf{k}'-\mathbf{q}j\beta\sigma'}
	d^{\phantom{\dag}}_{\mathbf{k}'j\beta\sigma'}.
\quad
\end{eqnarray}
In these equations, $\mu$ is the chemical potential, ${\cal N}$
is the number of unit cells in a bilayer sample, operators
$d^{\dag}_{\mathbf{k}i\alpha\sigma}$
and
$d^{\phantom{\dag}}_{\mathbf{k}i\alpha\sigma}$
are the creation and annihilation operators of the electrons with  momentum
$\mathbf{k}$
in the layer
$i$(=$1,\,2$),
in the sublattice
$\alpha$(=$A\,,B$)
with spin projection $\sigma$. The four-component operator-valued spinor
$\psi^{\dag}_{\mathbf{k}\sigma}$
is defined as
\begin{eqnarray}
\psi^{\dag}_{\mathbf{k}\sigma}
=
(d^{\dag}_{\mathbf{k}1A\sigma},\,d^{\dag}_{\mathbf{k}1B\sigma},\,
d^{\dag}_{\mathbf{k}2A\sigma},\,d^{\dag}_{\mathbf{k}2B\sigma}),
\end{eqnarray}
and the 4$\times$4 matrix
${\cal H}_{\mathbf{k}}$
equals
\begin{equation}
\label{H0}
{\cal H}_{\mathbf{k}}
=
\left(\begin{array}{cccc}
	{e\Phi}/{2}&-tf_{\mathbf{k}}&0&t_0\\
	-tf_{\mathbf{k}}^{*}&{e\Phi}/{2}&0&0\\
	0&0&-{e\Phi}/{2}&-tf_{\mathbf{k}}\\
	t_0&0&-tf_{\mathbf{k}}^{*}&-{e\Phi}/{2}
\end{array}\right),
\end{equation}
where $e$ is the electron charge, $\Phi$ is the bias voltage, and the
function
$f_{\mathbf{k}}$
is
\begin{equation}
f_{\mathbf{k}}
=
e^{i\mathbf{k}\bm{\delta}}
\left[1+e^{-i\mathbf{ka}_1}+e^{-i\mathbf{ka}_2}\right].
\end{equation}
Parameter
$t=2.7$\,eV
is the in-plane nearest-neighbor hopping amplitude,
$t_0=0.4$\,eV
is the out-of-plane hopping amplitude between nearest-neighbor sites in
positions $1A$ and $2B$. We choose the values of the hopping amplitudes $t$
and
$t_0$
in accordance with
Ref.~\onlinecite{bilayer_review2016}.

It is important that in our model the interaction function $V^{ij}_{\mathbf{q}}$ in
Eq.~\eqref{Hint} is not a bare Coulomb electron-electron repulsion.
It is a renormalized interaction, which accounts for many-body
screening effects. It will be evaluated below using the RPA. As for
electron-lattice coupling, it is ignored in our analysis.

We distinguish in the interaction Hamiltonian~\eqref{Hint}
the intra-layer and inter-layer couplings. This is done by introducing the layer indices in
$V^{ij}_{\mathbf{q}}$. The interaction can be represented as a 2$\times$2 matrix. In
such a matrix, the diagonal elements correspond to the intra-layer
interaction, while the off-diagonal elements correspond to the inter-layer one.

Solving the eigenvalue/eigenvector problem
${\cal H}_{\bf k} \Psi_{\bf k} = \varepsilon_{\bf k} \Psi_{\bf k}$
for
matrix~\eqref{H0}
we obtain the single-particle spectrum of AB-BLG. It consists of the four
bands
\begin{eqnarray}
\label{spec}
\varepsilon^{(1)}_{\mathbf{k}}
&=&
-\sqrt{
	t_{\mathbf{k}}^2
	+
	\frac{e^2\Phi^2}{4}
	+
	\frac{t_0^2}{2}
	+
	\sqrt{t_{\mathbf{k}}^2(e^2 \Phi^2+t_0^2) + \frac{t_0^4}{4}} },
\nonumber
\\
\varepsilon^{(2)}_{\mathbf{k}}
&=&
-\sqrt{
	t_{\mathbf{k}}^2
	+
	\frac{e^2\Phi^2}{4}
	+
	\frac{t_0^2}{2}
	-
	\sqrt{t_{\mathbf{k}}^2(e^2 \Phi^2+t_0^2) + \frac{t_0^4}{4}} },
\nonumber
\\
\varepsilon^{(3)}_{\mathbf{k}}
&=&
\sqrt{
	t_{\mathbf{k}}^2
	+
	\frac{e^2\Phi^2}{4}
	+
	\frac{t_0^2}{2}
	-
	\sqrt{t_{\mathbf{k}}^2(e^2 \Phi^2+t_0^2) + \frac{t_0^4}{4}} },
\nonumber
\\
\varepsilon^{(4)}_{\mathbf{k}}
&=&
\sqrt{
	t_{\mathbf{k}}^2
	+
	\frac{e^2\Phi^2}{4}
	+
	\frac{t_0^2}{2}
	+
	\sqrt{t_{\mathbf{k}}^2(e^2 \Phi^2+t_0^2) + \frac{t_0^4}{4}} },
\end{eqnarray}
where
$t_{\mathbf{k}}=t|f_{\mathbf{k}}|$.
When
$e\Phi =0$,
the spectrum near the Dirac points
$\mathbf{K}_1=(0, 4\pi/3a)$
and
$\mathbf{K}_2=-\mathbf{K}_1$
consists of four parabolic bands (two electron and two hole bands) with one
electron and one hole bands touching each other at Dirac points. At finite
$e \Phi$
a single-particle gap opens, and the AB-BLG becomes an insulator.

The bi-spinor wave functions
\begin{eqnarray}
\Psi_{{\bf k} }^{(S)}
=
\left(
	\Psi_{{\bf k} 1 A}^{(S)},\,
	\Psi_{{\bf k} 1 B}^{(S)},\,
	\Psi_{{\bf k} 2 A}^{(S)},\,
	\Psi_{{\bf k} 2 B}^{(S)}
\right),
\end{eqnarray}
corresponding to the eigenvalues
$\varepsilon^{(S)}_{\bf k}$,
$S=1, \ldots, 4$,
can be expressed analytically as well. However, the resultant formulas are
quite cumbersome. In what follows we will evaluate
$\Psi_{{\bf k} }^{(S)}$
numerically.

It is useful to introduce new electronic operators
$\gamma^{\dag}_{\mathbf{k}S\sigma}$
and
$\gamma^{\phantom{\dag}}_{\mathbf{k}S\sigma}$
according to
\begin{eqnarray}
\label{eq::dtopsi}
d^{\phantom{\dag}}_{\mathbf{k}i\alpha \sigma }
=
\sum_S
	\Psi^{(S)}_{\mathbf{k}i\alpha}
	\gamma^{\phantom{\dag}}_{\mathbf{k}S\sigma}.
\end{eqnarray}
Operator
$\gamma^{\dag}_{\mathbf{k}S\sigma}$
(operator
$\gamma^{\vphantom{\dag}}_{\mathbf{k}S\sigma}$)
creates (destroys) an electron in an eigenstate with quasi-momentum
${\bf k}$
in band $S$. In terms of these operators the single-particle Hamiltonian
reads
\begin{eqnarray}
H_0=\sum_{\mathbf{k}S\sigma}
	(\varepsilon^{(S)}_{\mathbf{k}} - \mu)
	\gamma^{\dag}_{\mathbf{k}S\sigma}
	\gamma^{\phantom{\dag}}_{\mathbf{k}S\sigma}\,.
\end{eqnarray}

\begin{figure*}
\centering
\includegraphics[width=0.32\textwidth]{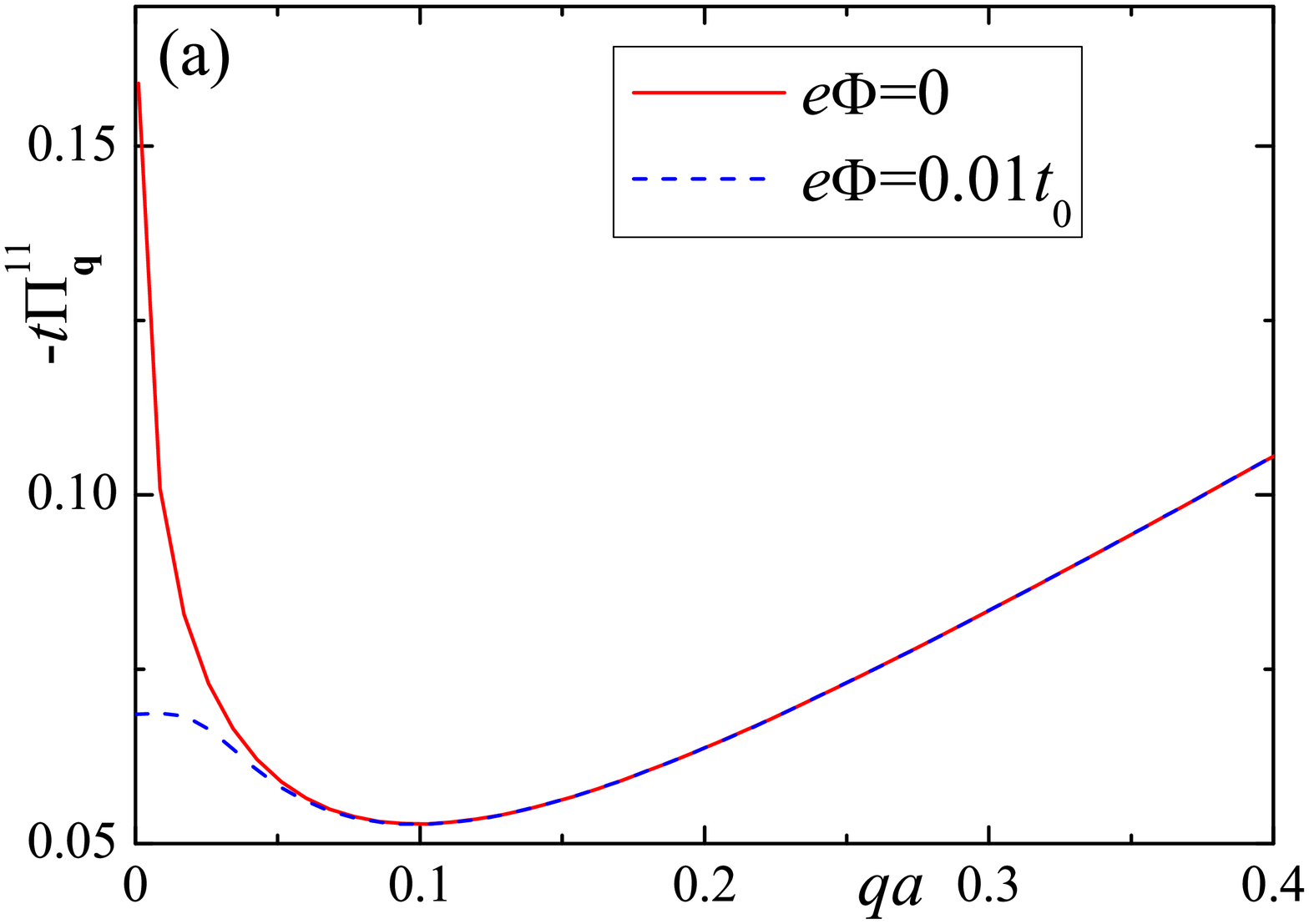}
\includegraphics[width=0.32\textwidth]{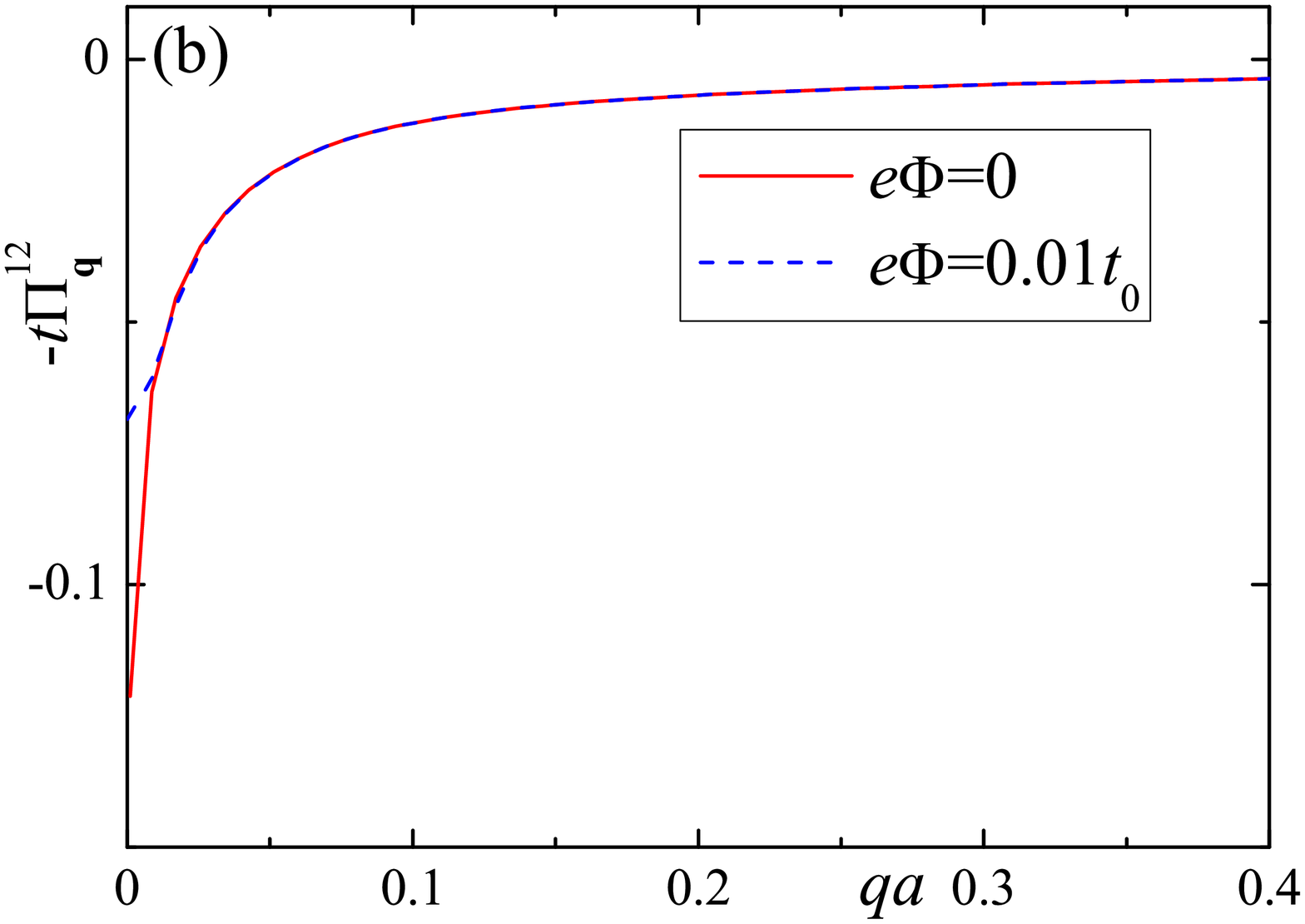}
\includegraphics[width=0.33\textwidth]{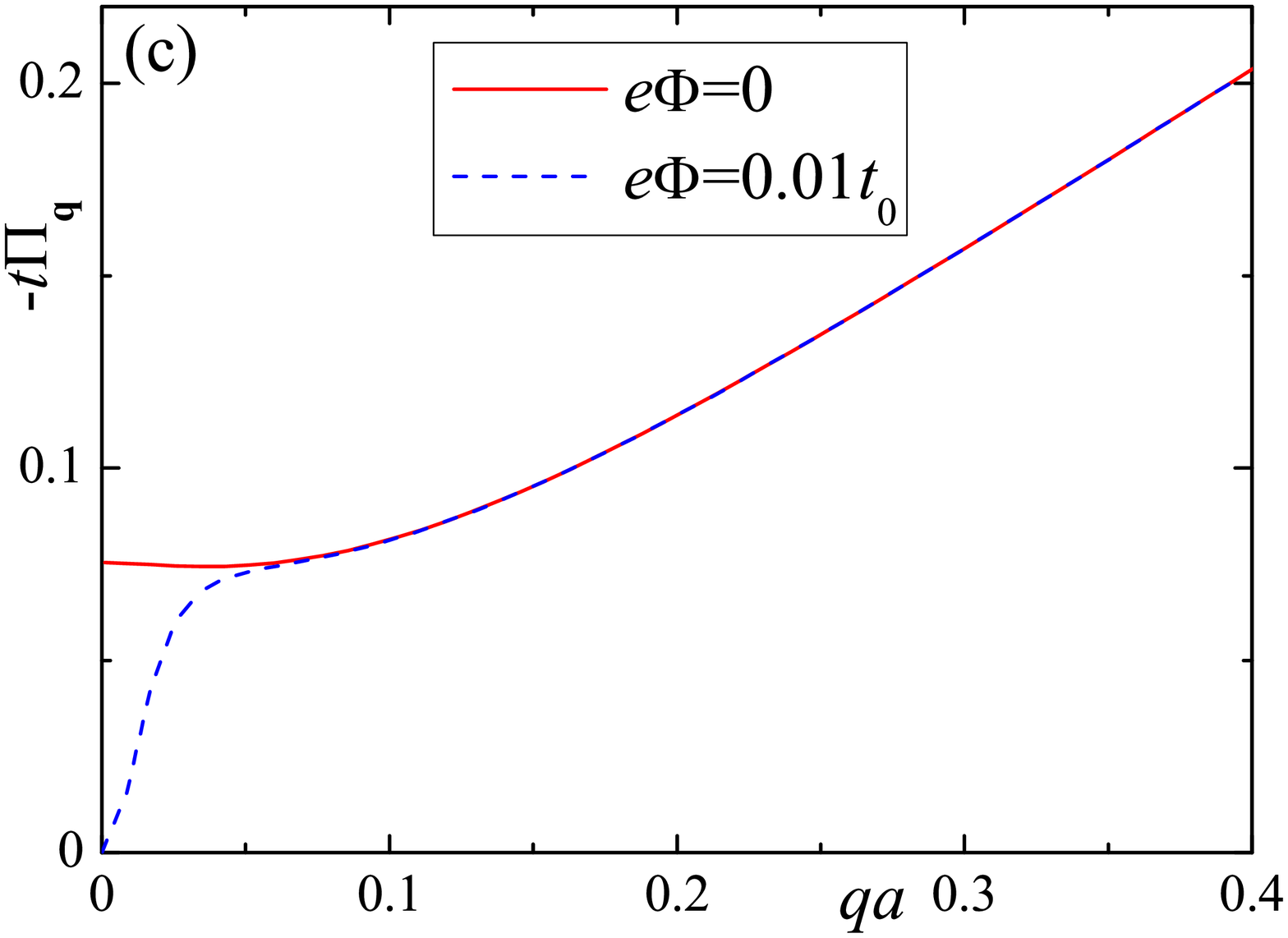}
\caption{
\label{FigPundoped}
The polarization operator components of the undoped AB-BLG, as functions of the
momentum
${\bf q}$,
for
$\mathbf{q}=q(1,\,0)$.
Panels~(a), (b), and~(c) show
$\Pi^{11}_{\mathbf{q}}$,
$\Pi^{12}_{\mathbf{q}}$,
and the total polarization operator
$\Pi_{\mathbf{q}}=\sum_{ij}\Pi^{ij}_{\mathbf{q}}$,
respectively. The curves in all panels are calculated for
$\mu=0$.
Solid (red) curves correspond to
$e\Phi=0$,
while dashed (blue) curves correspond to
$e\Phi=0.01t_0$.
}
\end{figure*}

\section{polarization operator and renormalized Coulomb potential for
undoped bilayer}
\label{sec::undoped_RPA}

Coulomb interaction in a solid experiences unavoidably strong
renormalization due to screening. As already mentioned, we assume that the
interaction function $V^{ij}_{\mathbf{q}}$ in
Hamiltonian~(\ref{Hint})
incorporates static screening effects. To calculate
$V^{ij}_{\mathbf{q}}$,
the RPA can be used. It is commonly believed that for graphene-based systems
the RPA is a more appropriate approach due to the larger degeneracy factor
$N_d=4$.

A key element of any RPA scheme is a polarization operator. During two
decades of the theoretical research on graphene numerous workers calculated
the polarization operator for both biased and unbiased AB-BLG (see, e.g.,
Refs.~\onlinecite{Pol1TwoBandChakrabortyPRB2007,
Pol2TwoBandBiasedChakrabortyPRB2010,Pol3TwoBandWanPRB2010,
Pol7TwoBandDasSarmaPRL2008,Pol4TwoBandDasSarmaPRB2010,
Pol5FourBandGamayunPRB2011,Pol6FourBandRossiPRB2012}).
In the most of these publications the effective two-band model of the
AB-BLG has been employed. In
Refs.~\onlinecite{Pol5FourBandGamayunPRB2011,Pol6FourBandRossiPRB2012}
the polarization operator is calculated in the framework of four-band model
using continuum approximation.

In this paper we numerically evaluate the static polarization operator
$\Pi^{ij}_{\mathbf{q}}$
for the four-band tight-binding model. Both intra-layer
($i=j$)
and inter-layer
($i \ne j$)
components will be determined.
This is to be contrasted with the majority
of the previous studies that considered the total polarization operator
$\Pi_{\mathbf{q}}=\sum_{ij}\Pi^{ij}_{\mathbf{q}}$
only.

The polarization operator of the undoped AB-BLG can be presented as a
$2\times2$
matrix. The elements of this matrix as functions of the transferred
momentum $\mathbf{q}$ reads~\cite{Pol6FourBandRossiPRB2012}
\begin{eqnarray}
\label{P}
\Pi^{ij}_{\mathbf{q}}&=&2\sum_{SS'}\int\!\frac{d^2\mathbf{k}}{v_{\rm BZ}}
\frac{n_{\rm F}\left(\varepsilon^{(S)}_{\mathbf{k}}\right)-n_{\rm F}\left(\varepsilon^{(S')}_{\mathbf{k}+\mathbf{q}}\right)}
{\varepsilon^{(S)}_{\mathbf{k}}-\varepsilon^{(S')}_{\mathbf{k}+\mathbf{q}}}\times\nonumber\\
&\times&\Big(\sum_{\alpha}\Psi^{(S)}_{\mathbf{k}i\alpha}\Psi^{(S')*}_{\mathbf{k}+\mathbf{q}i\alpha}\Big)
\Big(\sum_{\beta}\Psi^{(S)*}_{\mathbf{k}j\beta}\Psi^{(S')}_{\mathbf{k}+\mathbf{q}j\beta}\Big),
\end{eqnarray}
where
$S,\,S'=1,\dots,\,4$,
$v_{\rm BZ}=8\pi^2/(a^2\sqrt{3})$
is the Brillouin zone area, and
$n_{\rm F}(E)=[e^{(E-\mu)/T}+1]^{-1}$
is the Fermi function. We limit ourselves to zero temperature.
The results of the numerical calculations of $\Pi^{ij}_{\mathbf{q}}$
are shown in Fig.~\ref{FigPundoped}
for two different values of $e\Phi$
($e\Phi = 0$
and
$e\Phi = 0.01t_0$).

\begin{figure}[t]
\centering
\includegraphics[width=0.95\columnwidth]{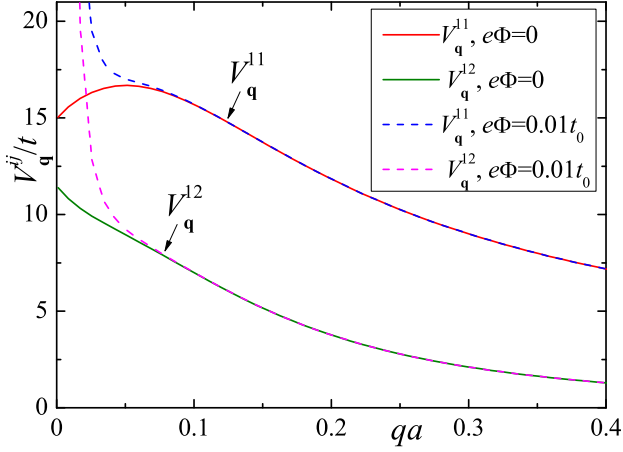}
\caption{
\label{FigVundoped}
The renormalized interaction components as functions of the momentum
$q$, at $\mathbf{q}=q(1,\,0)$,
calculated for the undoped AB-BLG
($\mu=0$)
and two values of the bias potential
$e\Phi$.
Diagonal components
$V^{11}_{\mathbf{q}} = V^{22}_{\mathbf{q}}$
are plotted as red and blue curves, off diagonal
$V^{12}_{\mathbf{q}} = V^{21}_{\mathbf{q}}$
components are plotted as green and magenta curves. The solid (dashed)
curves represent
$e\Phi=0$
case
($e\Phi=0.01t_0$
case). For all curves
$\epsilon=1$.
}
\end{figure}

Analyzing the numerical data we observe that
$\Pi^{11}_{\mathbf{q}}=\Pi^{22}_{\mathbf{q}}$,
which a manifestation of the charge-conjugation symmetry (see also
Appendix~\ref{append::charge_conj}).
Further, as long as
$q=|\mathbf{q}|$
is not too large,
$qa<1$,
the polarization operator is virtually independent of the direction of
$\mathbf{q}$.
From
Fig.~\ref{FigVundoped}
we see that the intra-layer components
$\Pi^{11}_{\mathbf{q}}$ and $\Pi^{22}_{\mathbf{q}}$
are always negative, while
$\Pi^{12}_{\mathbf{q}}$
is positive. For small $q$, the value of
$-\Pi^{11}_{\mathbf{q}}$
decreases with the increase of $q$. This decay is replaced by a linear
growth at larger $q$, which is similar to the behavior of the polarization
operator of the single-layer
graphene~\cite{InteractionsInGrapheneRevModPhys2012}.
The inter-layer polarization
$\Pi^{12}_{\mathbf{q}}$
monotonously decreases with $q$. Asymptotically, it behaves as
$1/q$
at
$qa>0.1$.

The renormalized Coulomb interaction can be expressed in the matrix form as
\begin{equation}
\label{VRPA}
\hat{V}_{\mathbf{q}}
=
\hat{V}^{(0)}_{\mathbf{q}}
\left( 1 - \hat{\Pi}_{\mathbf{q}}\hat{V}^{(0)}_{\mathbf{q}} \right)^{-1}.
\end{equation}
In this formula, the bare Coulomb interaction is a
$2\times2$ matrix
\begin{equation}
\label{V0}
V^{(0)}_{\mathbf{q}}=\frac{A}{q}\left(\begin{array}{cc}
1&e^{-qd}\\
e^{-qd}&1
\end{array}\right),\;\;
A=\frac{2\pi e^2}{{\cal S}_{\rm gr}\epsilon} \,,
\end{equation}
where
${\cal S}_{\rm gr}=a^2\sqrt{3}/2$
is the area of the graphene unit cell, and $\epsilon$ is the dielectric
constant of the media surrounding the graphene sample. Thus, we obtain
\begin{widetext}
\begin{eqnarray}
V^{11}_{\mathbf{q}}=V^{22}_{\mathbf{q}}&=&A\frac{1-\frac{A}{q}\Pi^{22}_{\mathbf{q}}[1-e^{-2qd}]}
{q-A(\Pi^{11}_{\mathbf{q}}+\Pi^{22}_{\mathbf{q}}+2e^{-qd}\Pi^{12}_{\mathbf{q}})+
\frac{A^2}{q}[\Pi^{11}_{\mathbf{q}}\Pi^{22}_{\mathbf{q}}-(\Pi^{12}_{\mathbf{q}})^2][1-e^{-2qd}]},
\label{V11}
\\
V^{12}_{\mathbf{q}}=V^{21}_{\mathbf{q}}&=&A\frac{e^{-qd}+\frac{A}{q}\Pi^{12}_{\mathbf{q}}[1-e^{-2qd}]}
{q-A(\Pi^{11}_{\mathbf{q}}+\Pi^{22}_{\mathbf{q}}+2e^{-qd}\Pi^{12}_{\mathbf{q}})+
\frac{A^2}{q}[\Pi^{11}_{\mathbf{q}}\Pi^{22}_{\mathbf{q}}-(\Pi^{12}_{\mathbf{q}})^2][1-e^{-2qd}]}.
\label{V12}
\end{eqnarray}
\end{widetext}
Similar results can be found in the literature on the Coulomb drag in
two-dimensional systems, see, for example,
Refs.~\onlinecite{coulomb_drag1995theor}
and~\onlinecite{coulomb_drag1995theor2}.

At zero bias we have
$\Pi^{11}_\mathbf{q} + \Pi^{12}_\mathbf{q} \neq0$
at
$q\to0$
(which is consistent with the results obtained in
Ref.~\onlinecite{Pol7TwoBandDasSarmaPRL2008}).
Thus, the matrix
$\hat{V}_{\mathbf{q}}$
is regular at
$q\to0$.
In other words, the screened Coulomb potential is finite at
$q=0$,
which agrees with a general expectation that finite density of states at
the Fermi energy leads to the suppression of the long-range Coulomb
interaction.

When $e\Phi\neq0$,
the single-electron spectrum acquires a gap that affects the low-$q$
screening. Indeed, in this regime $(\Pi^{11}_{\bf q} + \Pi^{12}_{\bf q}) \vert_{{\bf q} =0} = 0$,
thus, the matrix $\hat{V}_{\mathbf{q}}$
is singular at $q=0$.
This singularity indicates that in the insulating state of the biased AB-BLG
the long-range interaction cannot be completely screened
and the resultant Coulomb potential behaves as
$V^{ij}_{\mathbf{q}}\propto1/q$
at small $q$. However, such a behavior persists for small momenta only.
Additional details can be learned from
Fig.~\ref{FigVundoped}
where numerically calculated
$V^{ij}_{\mathbf{q}}$
is plotted for
$e\Phi=0$
and
$e\Phi=0.01t_0$.

Concluding this section, we would like to make the following observation.
If $d\to 0$,
then
Eqs.~(\ref{V11}) and~(\ref{V12})
are replaced by one simple formula
\begin{eqnarray}
\label{eq::V_ij_simple}
V^{ij}_{\mathbf{q}}
=
\frac{A}{q-A\Pi_{\mathbf{q}}}.
\end{eqnarray}
The right-hand side of this expression is independent of $i$ and $j$. In
other words, such an approximation implies that the inter-layer and
intra-layer interactions are identical. In the literature,
theoretical results essentially similar to
Eq.~(\ref{eq::V_ij_simple})
are not uncommon (see, for example,
Refs.~\onlinecite{Pol1TwoBandChakrabortyPRB2007,
Pol2TwoBandBiasedChakrabortyPRB2010, Pol7TwoBandDasSarmaPRL2008},
to name a few). Unfortunately, the reliability of this approximation is not
clear: our numerical data suggests that
formula~(\ref{eq::V_ij_simple})
is a rather crude simplification that is poorly applicable even in the limit of
small $q$. More details can be found in
Appendix~\ref{append::V_ij}.

%

\section{Spin-density wave state}
\label{sec::MF_SDW}

The computed renormalized Coulomb interaction can be applied to the
study of the AB-BLG ordered states.
We characterize the SDW by the following expectation value
\begin{equation}
\label{etaSDW}
\eta^{\text{SDW}}_{\mathbf{k}}
=
\left\langle
	\gamma^{\dag}_{\mathbf{k}3\bar{\sigma}}
	\gamma^{\phantom{\dag}}_{\mathbf{k}2\sigma}
\right\rangle,
\end{equation}
which we assumed to be independent of $\sigma$ (the bar over $\sigma$ means
not $\sigma$). This relation implies that in our SDW state a hole in the
band
$S=2$
is coupled to an electron with opposite spin in the band
$S=3$.

Equation~\eqref{eq::dtopsi}
allows us to express
$H_{\text{int}}$
in terms of the band operators
$\gamma^{\dag}_{\mathbf{k}S\sigma}$,
$\gamma^{\phantom{\dag}}_{\mathbf{k}S\sigma}$.
Keeping only the terms relevant to the SDW pairing, one derives
\begin{eqnarray}
\label{HintPsi}
H_{\text{int}}
&=&
-\frac{1}{2\cal N}
\sum_{\mathbf{kk}'\sigma}
\left(
	\gamma^{\dag}_{\mathbf{k}2\sigma}
	\gamma^{\phantom{\dag}}_{\mathbf{k}3\bar{\sigma}}
	\Gamma^{(1)}_{\mathbf{kk}'}
	\gamma^{\dag}_{\mathbf{k}'3\bar{\sigma}}
	\gamma^{\phantom{\dag}}_{\mathbf{k}'2\sigma}
\right.
+
\nonumber
\\
&&
\left.
	\gamma^{\dag}_{\mathbf{k}2\sigma}
	\gamma^{\phantom{\dag}}_{\mathbf{k}3\bar{\sigma}}
	\Gamma^{(2)}_{\mathbf{kk}'}
	\gamma^{\dag}_{\mathbf{k}'2\bar{\sigma}}
	\gamma^{\phantom{\dag}}_{\mathbf{k}'3\sigma}
	+
	{\rm H.c.}
\right),
\end{eqnarray}
where
\begin{eqnarray}
\label{Gamma12}
\Gamma^{(1)}_{\mathbf{kk}'}
&=&
\sum_{ij}\!
	\Big(\!
		\sum_{\alpha}
			\Psi^{(2)*}_{\mathbf{k}i\alpha}
			\Psi^{(2)}_{\mathbf{k}'i\alpha}
	\Big)\!
	V^{ij}_{\mathbf{k}-\mathbf{k}'}\!
	\Big(\!
		\sum_{\beta}
			\Psi^{(3)}_{\mathbf{k}j\beta}
			\Psi^{(3)*}_{\mathbf{k}'j\beta}
	\Big),
\;\;\;\;
\nonumber\\
\Gamma^{(2)}_{\mathbf{kk}'}
&=&
\sum_{ij}\!
	\Big(\!
		\sum_{\alpha}
			\Psi^{(2)*}_{\mathbf{k}i\alpha}
			\Psi^{(3)}_{\mathbf{k}'i\alpha}
	\Big)\!
	V^{ij}_{\mathbf{k}-\mathbf{k}'}\!
	\Big(\!
		\sum_{\beta}
			\Psi^{(3)}_{\mathbf{k}j\beta}
			\Psi^{(2)*}_{\mathbf{k}'j\beta}
	\Big).\;\;\;\;
\end{eqnarray}
Note that
Eq.~(\ref{HintPsi})
ignores retardation effects in screening physics, implying that the
screening is instantaneous. The validity of this approximation will be
discussed in
subsection~\ref{subsec::RPA_issues}.

Introducing the SDW order parameter as
\begin{equation}
\label{deltaSDW}
\Delta^{\text{SDW}}_{\mathbf{k}}
=
\frac{1}{\cal N}\sum_{\mathbf{k}'}
	\left(
		\Gamma^{(1)}_{\mathbf{kk}'}\eta^{\text{SDW}}_{\mathbf{k}'}
		+
		\Gamma^{(2)}_{\mathbf{kk}'}\eta^{\text{SDW}*}_{\mathbf{k}'}
	\right)\,,
\end{equation}
and performing the standard mean-field decoupling scheme in
Eq.~\eqref{HintPsi},
we obtain the mean-field Hamiltonian, which allows us to calculate the
grand potential $\Omega$. Minimization of $\Omega$ gives the following
equation for the SDW order parameter:
\begin{equation}
\label{DeltaSDWeq}
\Delta^{\text{SDW}}_{\mathbf{k}}
=
\int\!\frac{d^2\mathbf{k}'}{v_{\rm BZ}}\,
	\frac{\Gamma^{(1)}_{\mathbf{kk}'}\Delta^{\text{SDW}}_{\mathbf{k}'}
	+
	\Gamma^{(2)}_{\mathbf{kk}'}\Delta^{\text{SDW}*}_{\mathbf{k}'}}
	{2\sqrt{
		\big[\varepsilon^{(3)}_{\mathbf{k}'}\big]^2
		+
		|\Delta^{\text{SDW}}_{\mathbf{k}'}|^2
		}
	}.
\end{equation}
Let us consider first the case of
$e\Phi=0$.
We do not solve the integral
equation~\eqref{DeltaSDWeq} directly.
Instead, we perform a transparent and physically motivated approximate evaluation of $\Delta^{\text{SDW}}_{\mathbf{k}}$.
First, we observe that the main contribution to the integral in the right-hand
side of Eq.~\eqref{DeltaSDWeq}
comes from momenta $\mathbf{k}'$
near the Dirac points
$\mathbf{K}_{\xi}$
($\xi=1,\,2$).
Thus, it is necessary to know the behavior of
$\Gamma^{(1)}_{\mathbf{kk}'}$
and
$\Gamma^{(2)}_{\mathbf{kk}'}$
with momenta
$\mathbf{k}$
and
$\mathbf{k}'$
close to
$\mathbf{K}_{\xi}$.
It is possible to show that, for
$e\Phi=0$,
the wave functions
$\Psi^{(2,3)}_{\mathbf{k}\alpha}$
near the Dirac point
$\mathbf{K}_{\xi}$
are
\begin{eqnarray}
\label{Phi23V0}
\Psi^{(2)}_{\mathbf{K}_{\xi}+\mathbf{p}i\alpha}
&=&
\frac{1}{\sqrt{2}}
\left(
	\begin{array}{c}
		0\\
		e^{-i[\frac{\pi}{2}-(-1)^{\xi}\phi_{\mathbf{p}}]}\\
		e^{i[\frac{\pi}{2}-(-1)^{\xi}\phi_{\mathbf{p}}]}\\
		0
	\end{array}
\right),
\nonumber\\
\Psi^{(3)}_{\mathbf{K}_{\xi}+\mathbf{p}i\alpha}
&=&
\frac{1}{\sqrt{2}}
\left(
	\begin{array}{c}
		0\\
		-e^{-i[\frac{\pi}{2}-(-1)^{\xi}\phi_{\mathbf{p}}]}\\
		e^{i[\frac{\pi}{2}-(-1)^{\xi}\phi_{\mathbf{p}}]}\\
		0
	\end{array}
\right),
\end{eqnarray}
where
$\phi_{\mathbf{p}}$
is the polar angle of the vector
$\mathbf{p}\to 0$.
Substituting these equations in
formulas~\eqref{Gamma12},
we approximate
$\Gamma^{(1,2)}_{\mathbf{K}_{\xi}+\mathbf{p}\mathbf{K}_{\xi}+\mathbf{p}'}$
for small
$|{\bf p}|$
and
$|{\bf p}'|$
as
\begin{eqnarray}
\label{Gamma12Dirac}
\Tilde \Gamma^{(1,2)}_{\mathbf{p}\mathbf{p}'}
&\approx&
\frac12\left\{
	V^{11}_{\mathbf{p} - \mathbf{p}'}
	\pm
	V^{12}_{\mathbf{p} - \mathbf{p}'}
	\cos\left[2(\phi_{\mathbf{p}}-\phi_{\mathbf{p}'})\right]
\right\}.
\end{eqnarray}
Here and below the tilde over a function of momentum indicates that the
momentum is measured from the Dirac point
$\mathbf{K}_{\xi}$.
The approximate quantities
$\Tilde \Gamma^{(1,2)}_{{\bf p} {\bf p}'}$
are the same in both valleys, consequently, dependence on $\xi$ is
suppressed. Additionally,
Eq.~(\ref{Gamma12Dirac})
implies that
$\Tilde\Gamma^{(1,2)}_{\mathbf{pp}'}$
are real functions of momenta
$\mathbf{p}$
and
$\mathbf{p}'$
when
$\mathbf{p}$,
$\mathbf{p}'$
are close to a Dirac point. Therefore, one can expect that the SDW order
parameter is also a real function of
$\mathbf{p}$.

To estimate the SDW order parameter, we assume that
$\Delta^{\text{SDW}}_{\mathbf{k}}$
is a step function of the momentum inside some region near each Dirac point, that is,
\begin{equation}
\label{ansatz}
\Delta^{\text{SDW}}_{\mathbf{K}_{\xi}+\mathbf{p}}
=
\left\{\begin{array}{cl}
\Delta^{\text{SDW}}&,\;\;|\mathbf{p}|<K_0\\
0&,\;\;|\mathbf{p}|>K_0
\end{array}\right.\,,
\end{equation}
where the cutoff momentum
$K_0$
is chosen such that the regions corresponding to different Dirac points do
not intersect. Below we neglect coupling of the order parameters from
different valleys and assume that
$\mathbf{k}$
and
$\mathbf{k}'$
in
Eq.~\eqref{DeltaSDWeq}
lie in the same valley. Taking
$\mathbf{k}=\mathbf{K_{\xi}}$
and using the
ansatz~\eqref{ansatz},
one derives the equation for
$\Delta^{\text{SDW}}$
\begin{equation}
\label{DeltaSDWtrans}
\int\limits_{|\mathbf{p}|<K_0}\!\!\!\!
	\frac{d^2\mathbf{p}}{v_{\rm BZ}}\,
	\frac{
		\Tilde \Gamma^{(1)}_{0\mathbf{p}}
		+
		\Tilde \Gamma^{(2)}_{0\mathbf{p}}
	}
	{2\sqrt{
		\big[\Tilde \varepsilon^{(3)}_{\mathbf{p}}\big]^2
		+
		\big(\Delta^{\text{SDW}}\big)^2
		}
	}=1\,.
\end{equation}
We solve this equation numerically, taking
$K_0$
by its maximum possible value
$K_0=2\pi/(3a)$.
We choose $\epsilon=1$. Other parameters are fixed as explained above. In so doing, we obtain
$\Delta^{\text{SDW}}=0.0018t=4.9$\,meV.
This result is in agreement with experimentally available data,
Ref.~\onlinecite{GapExp3PRB2012},
where the measured transport gap in the Bernal bilayer graphene, which is
twice of the order parameter, is equal to
$\Delta_{\text{tr}}=8$\,meV.

In our approach the SDW order arises due to the long-range Coulomb interaction. Thus, the result is sensitive to the value of dielectric constant: if $\epsilon$
is increased, the order parameter decreases. For example, for
$\epsilon = 5$,
we find that
$\Delta^{\rm SDW} = 0.15$\,meV,
which is about thirty times smaller than the value of
$\Delta^{\rm SDW}$
at
$\epsilon = 1$.

\begin{figure}[t]
\centering
\includegraphics[width=1\columnwidth]{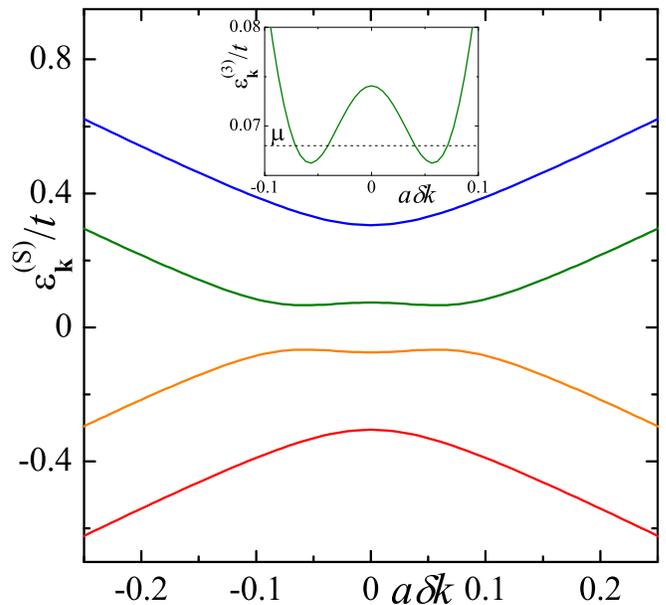}
\caption{
\label{FigBand}
Band structure of the biased bilayer graphene near Dirac point,
calculated at
$e\Phi=0.5t_0$.
Four bands
$S=1,\ldots, 4$
are plotted as functions of the deviation of momentum from the Dirac point
$\delta k$.
Inset shows the fine structure of the band
$\varepsilon^{(3)}_{\mathbf{k}}$
close to Dirac point. Horizontal dashed line is the position of the
chemical potential $\mu$.
}
\end{figure}
\begin{figure*}[t]
\centering
\includegraphics[width=0.32\textwidth]{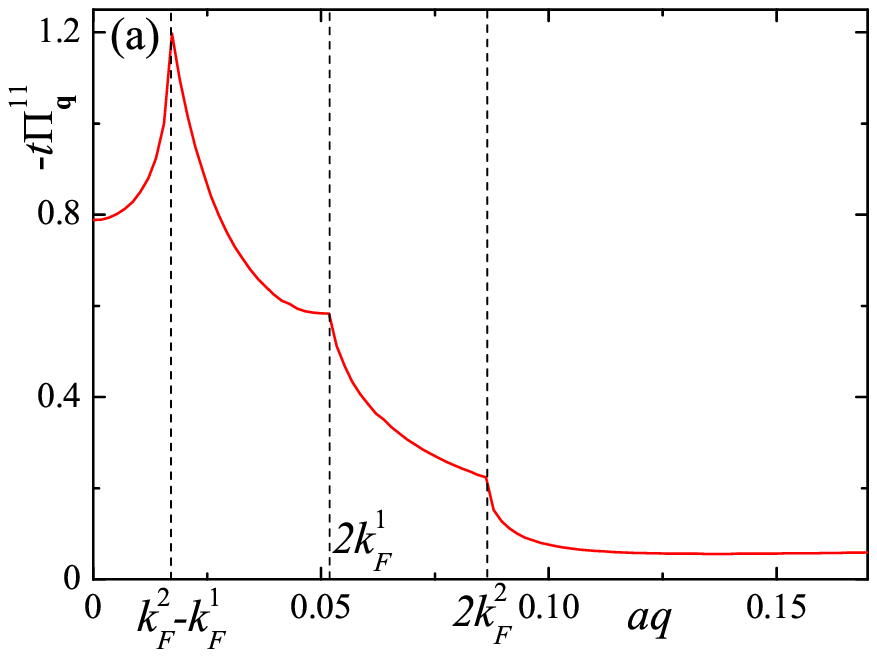}
\includegraphics[width=0.32\textwidth]{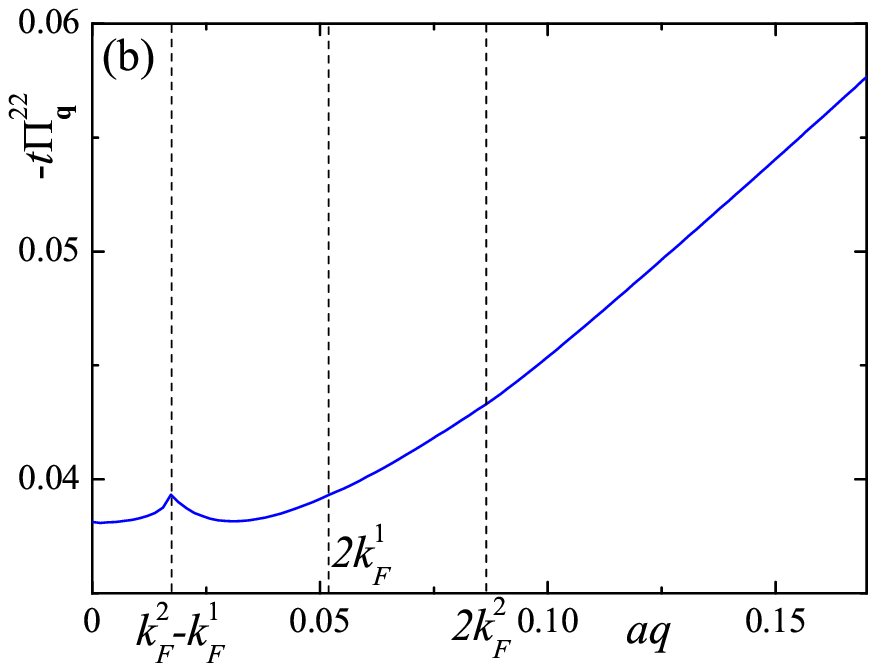}
\includegraphics[width=0.33\textwidth]{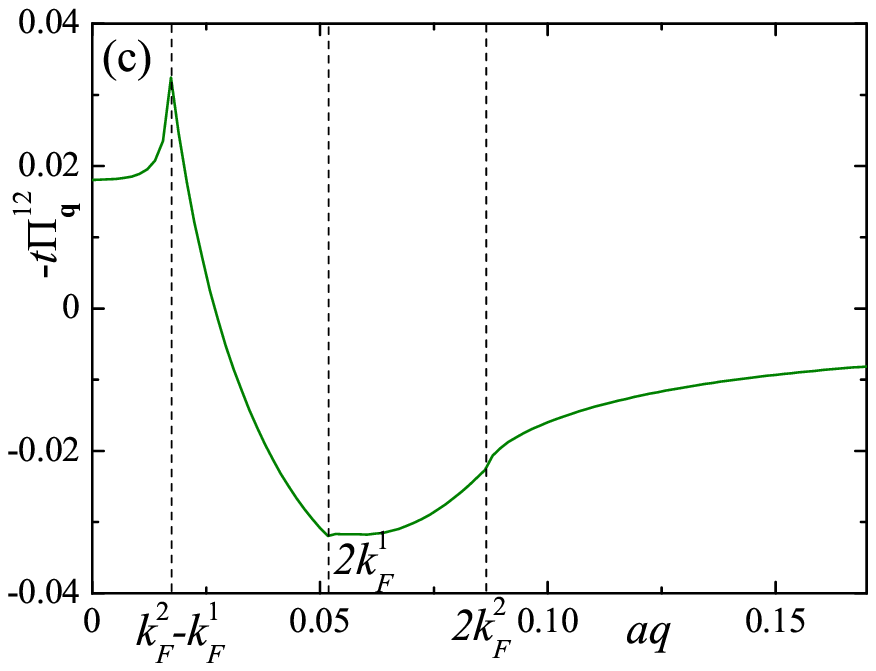}
\caption{
\label{FigP}
The polarization operator components for biased doped AB-BLG, as functions
of the momentum
${\bf q}$,
for
$\mathbf{q}=q(1,\,0)$.
Panels~(a), (b), and~(c) show
$\Pi^{11}_{\mathbf{q}}$,
$\Pi^{22}_{\mathbf{q}}$,
and
$\Pi^{12}_{\mathbf{q}}$,
respectively. The curves in all panels are calculated for
$e\Phi =0.3t_0$
and
$\mu=0.021t$.
}
\end{figure*}

Consider now the case of
$e\Phi \neq0$.
At finite bias, the gap between bands $2$ and $3$ arises even in the
single-particle approximation. Therefore, one can expect that the bias voltage
destroys the SDW ordering. Indeed, if the gap is open,
the denominator in
Eq.~\eqref{DeltaSDWeq}
never reaches zero even in the limit of
$\Delta^{\text{SDW}} \to0$.
As a result, we obtain from $\Delta^{\text{SDW}} \to0$
the following criterion for the existence of the
SDW ordering at finite bias voltage
\begin{equation}
\label{SDWcriterion}
\int\limits_{|\mathbf{p}|<K_0}\!\!\!\!
	\frac{d^2\mathbf{p}}{v_{\rm BZ}}\,
	\frac{
		\Tilde \Gamma^{(1)}_{\mathbf{p}'\mathbf{p}}
		+
		\Tilde \Gamma^{(2)}_{\mathbf{p}'\mathbf{p}}
	}{2\Tilde \varepsilon^{(3)}_{\mathbf{p}}}>1\,,
\end{equation}
where the momentum
$\mathbf{p}'$
is to be chosen to maximize the integral.
The values
$V^{ij}_{\mathbf{q}}$
for
$e\Phi =0$
and
$e\Phi \neq0$
almost coincide at larger
${\bf q}$
(see
Fig.~\ref{FigVundoped})
and we assume that $e\Phi$ is small enough.
Then, in Eq.~\eqref{SDWcriterion} we can use the functions
$\Gamma^{(1,2)}_{\mathbf{kk}'}$
calculated at
$e\Phi=0$
(divergence of
$V^{ij}_{\mathbf{q}}$
for
$e\Phi\neq0$
at
$q\to0$
is an integrable one). In this case, one can take
$\mathbf{k}=\mathbf{K}_{\xi}$,
or, equivalently,
$\mathbf{p}'=0$
in
Eq.~(\ref{SDWcriterion}).

Numerical analysis shows that SDW ordering is completely suppressed for
$e\Phi > e\Phi_c$,
where the critical bias value is found to be
$e\Phi_c=0.0038t=0.025t_0=10$\,meV
at
$\epsilon=1$.
Thus, we obtain quite natural result that the critical bias voltage
$e\Phi_c$
is of the order of
$\Delta^{\text{SDW}}$
calculated at
$e\Phi=0$.

\section{polarization operator and renormalized Coulomb potential for doped
bilayer}
\label{sec::doped_RPA}

The non-superconducting ordered state (for example, the SDW discussed
above, or a similar phase) is expected to dominate any superconducting
state in pristine graphene-based systems. Indeed, experimentally measured
energy scales associated with non-superconducting ordered phases are in the
range of several meV (see, for example,
Refs.~\onlinecite{Mayorov2011,GapExp2NatureNano2012,GapExp3PRB2012,
Freitag20122053}),
while the relevant superconducting energy is several orders of magnitude
lower~\cite{SCBLGNature2022}.
Consequently, it is necessary to suppress a non-superconducting order
parameter to make a superconducting transition possible.

The suppression of the SDW by the bias voltage, considered in
Sec.~\ref{sec::MF_SDW},
is not suitable since it leads to a change from the SDW insulator to the
band insulator. A more convenient approach is doping. Doping destroys the
SDW ordering, replacing it by a metal with a well-developed Fermi surface.

The presence of a Fermi surface drastically changes the screening
properties of AB-BLG. To account for these, we present here the results of
our numerical calculations of the polarization operator and renormalized
Coulomb potential of the doped and biased bilayer graphene. We consider an
electron doping and assume that under doping only the band
$S=3$
crosses the Fermi level $\mu$, while the band
$S=4$
remains empty. It is also assumed that the following restriction on the
chemical potential is met:
$\mu_{\text{min}}<\mu<\mu_{\text{max}}$,
where
$\mu_{\text{min}}=e\Phi t_0/(2\sqrt{e^2\Phi^2+t_0^2})$
and
$\mu_{\text{max}}=e\Phi/2$.
In this case the Fermi surface consists of four approximately circular
pockets. A pair of these, with Fermi momenta
$k_{\rm F}^{(1)}$
and
$k_{\rm F}^{(2)}$,
are centered at Dirac point
$\mathbf{K}_1$.
An identical pair is centered at
$\mathbf{K}_2$.
Using
Eq.~\eqref{spec}
and linear expansion of
$|t_{{\bf K}_\xi + \mathbf{k}}| \approx v_{\rm F} k$, 
where
$v_{\rm F} = \sqrt{3}t a/2$
is the graphene Fermi velocity, we derive an expression for the Fermi
momenta
$k_{\rm F}^{(1)}$
and
$k_{\rm F}^{(2)}$
\begin{equation}
\label{kFs}
k^{(1,2)}_{\rm F}\!\!
=\!
\frac{1}{v_{\rm F}}\!\sqrt{
	\frac{e^2\Phi^2}{4}\!
	+\!
	\mu^2
	\mp
	\sqrt{e^2\Phi^2\mu^2\!
		+\!
		\left(\!\mu^2 \! - \! \frac{e^2\Phi^2}{4}\! \right)\!t_0^2}
}.
\end{equation}
Each inner Fermi surface is hole-like, while outer one is electron-like.
Absolute values of the Fermi velocities at each Fermi surface are equal to
($s=1,\,2$)
\begin{equation}
\label{vFs}
v_{\rm F}^{(s)}
=
\frac{v_{\rm F}^{2}k_{\rm F}^{(s)}}{\mu}
\left|
	1
	-
	\frac{e^2\Phi^2+t_0^2}
		{\sqrt{4\big[v_{\rm F}k_{\rm F}^{(s)}\big]^2
		\big(e^2\Phi^2+t_0^2\big)+t_0^4}}
\right|.
\end{equation}
When
$\mu\to\mu_{\text{min}}$,
these velocities vanish,
$v_{\rm F}^{(s)}\to0$,
and the density of states at the Fermi level diverges. Band structure near
the Dirac point
$\mathbf{K}_1$
and typical position of the chemical potential are plotted in
Fig.~\ref{FigBand}.
The electron concentration (per one site ) is a function of $\mu$ and can
be expressed as
\begin{equation}
\label{doping}
x
=
\frac12
\int\frac{d^2\mathbf{k}}{v_{\text{BZ}}}
	\,\Theta\!\left(\mu-\varepsilon_{\mathbf{k}}^{(3)}\right)
\approx
\frac{\pi}{v_{\rm BZ}}\left[(k^{(2)}_{\rm F})^2-(k^{(1)}_{\rm F})^2\right],
\end{equation}
where
$\Theta(E)$
is the Heaviside step function.

The numerical analysis of
Eq.~\eqref{P}
shows that the doping substantially modifies the polarization operator at
small $q$ and the change comes mainly from the intraband term (term with
$S=S'=3$)
in
Eq.~\eqref{P},
which is zero if
$\mu=0$.
The bias voltage breaks the symmetry between graphene layers. As a result
extra charge introduced by the doping accumulates mainly, say, in layer
$1$. Thus, we have
$\Pi^{11}_{\mathbf{q}}\neq\Pi^{22}_{\mathbf{q}}$.
It turns out that
$|\Pi^{11}_{\mathbf{q}}|\gg|\Pi^{22}_{\mathbf{q}}|
\sim
|\Pi^{12}_{\mathbf{q}}|$
at small $q$, that is, the screening in layer~$1$ is much greater than in
layer~$2$. The dependencies of
$\Pi^{11}_{\mathbf{q}}$,
$\Pi^{22}_{\mathbf{q}}$,
and
$\Pi^{12}_{\mathbf{q}}$
on $q$ are shown in
Fig.~\ref{FigP}.
We clearly see three Kohn anomalies located at momenta
$q=k_{\rm F}^{(2)}-k_{\rm F}^{(1)}$,
$q=2k_{\rm F}^{(1)}$,
and
$q=2k_{\rm F}^{(2)}$.
Under doping, the value of
$\Pi^{12}_{\mathbf{q}}$
is negative at small $q$ for a definite doping level; in this case it changes
sign at some value of $q$. The polarization component
$\Pi^{11}_{\mathbf{q}}$ is the main contributor to the total polarization
$\Pi_{\mathbf{q}}$. The dependence of $\Pi_{\mathbf{q}}$ on $q$ computed in this work is consistent with the results obtained in the
framework of four band continuum model in
Ref.~\onlinecite{Pol6FourBandRossiPRB2012}.

\begin{figure}[t]
\centering
\includegraphics[width=0.95\columnwidth]{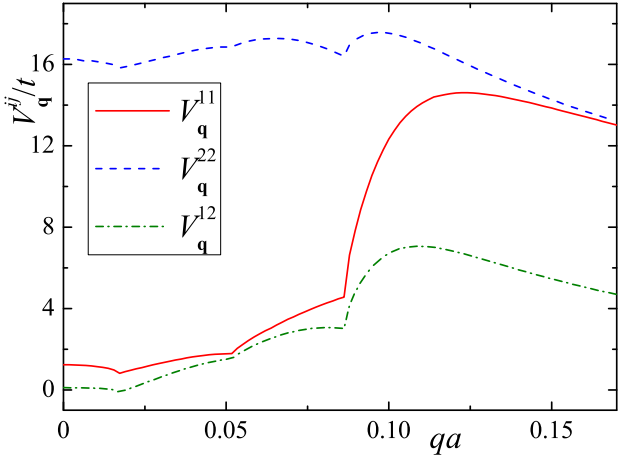}
\caption{
\label{FigV}
The renormalized interaction components at finite doping as functions of
the momentum
${\bf q}$,
for
$\mathbf{q}=q(1,\,0)$.
The curves are calculated at
$e\Phi=0.3t_0$
and
$\mu=0.0021t$.
Components
$V^{11}_{\mathbf{q}}$,
$V^{22}_{\mathbf{q}}$,
and
$V^{12}_{\mathbf{q}}$
are plotted as (red) solid curve, (blue) dashed curve, and (green)
dash-dotted curve, respectively. All curves are computed at $\epsilon=1$.
}
\end{figure}

The typical dependence of
$V^{ij}_{\mathbf{q}}$
on $q$ (for
$\epsilon=1$)
at finite doping and bias voltage is shown in
Fig.~\ref{FigV}.
In this regime
$\Pi^{11}_{\mathbf{q}}\neq\Pi^{22}_{\mathbf{q}}$,
consequently,
$V^{11}_{\mathbf{q}}\neq V^{22}_{\mathbf{q}}$.
When
$q\lesssim2k_{\rm F}^{(2)}$,
we have
$V^{11}_{\mathbf{q}}\ll V^{22}_{\mathbf{q}}$.
At small $q$ the screening  in the layer~2 is the weakest, thus,
the interaction inside this layer is the strongest. The screening effects
of the carriers introduced by doping become less important for larger $q$,
where $V^{11}_{\bf q}$ and
$V^{22}_{\bf q}$
are of the same order.

The important feature of the curves shown in
Fig.~\ref{FigV}
is that, when
$q\lesssim2k_{\rm F}^{(2)}$,
the interaction $V^{ij}_{\mathbf{q}}$
increases with $q$. As we will prove in the next section, such a
behavior is sufficient to stabilize a triplet superconducting state.

\section{Triplet superconductivity}
\label{sec::superconducting_mech}

The following consideration of superconductivity in biased and doped bilayer
graphene assumes that the bias voltage $e\Phi$ exceeds the critical value
$e\Phi_c$, thus, the SDW state is suppressed. The considered here type of the
superconductivity arises due to Coulomb interaction.
In contrast to usual BCS $s$-wave superconductivity, this phase exists
only in the $p$-wave channel, as it will be discussed below.

To derive the mean-field form of the model, we rewrite the interaction
Hamiltonian~\eqref{Hint} in the form
\begin{equation}
\label{HintSC}
H_{\text{int}}
\!=\!
\frac{1}{2\cal N}\!
\sum_{\mathbf{kk}' \sigma\sigma' \atop ij\alpha\beta}
	d^{\dag}_{\mathbf{k}i\alpha\sigma}
	d^{\dag}_{-\mathbf{k}j\beta\sigma'}
	V^{ij}_{\mathbf{k}-\mathbf{k}'}
	d^{\phantom{\dag}}_{-\mathbf{k}'j\beta\sigma'}
	d^{\phantom{\dag}}_{\mathbf{k}'i\alpha\sigma}\,,
\end{equation}
where all contributions unimportant for the superconductivity are omitted.
Substituting
Eq.~\eqref{eq::dtopsi}
in the formula above and keeping only terms with
$S=3$,
one obtains
\begin{equation}
\label{HintSCpsi}
H_{\text{int}}
=
\frac{1}{2\cal N}
\sum_{\mathbf{kk}' \sigma\sigma' \atop ij\alpha\beta}
	\gamma^{\dag}_{\mathbf{k}3\sigma}
	\gamma^{\dag}_{-\mathbf{k}3\sigma'}
	\Gamma_{\mathbf{kk}'}^{\rm SC}
	\gamma^{\phantom{\dag}}_{-\mathbf{k}'3\sigma'}
	\gamma^{\phantom{\dag}}_{\mathbf{k}'3\sigma}\,,
\end{equation}
where
\begin{equation}
\label{Gamma}
\Gamma_{\mathbf{kk}'}^{\rm SC}
=
\sum_{ij}\Big(
	\sum_{\alpha}
		\Psi^{(3)*}_{\mathbf{k}i\alpha}
		\Psi^{(3)}_{\mathbf{k}'i\alpha}
	\Big)
	V^{ij}_{\mathbf{k}-\mathbf{k}'}
	\Big(
	\sum_{\beta}
		\Psi^{(3)*}_{-\mathbf{k}j\beta}
		\Psi^{(3)}_{-\mathbf{k}'j\beta}
	\Big).
\end{equation}
The role of
$\Gamma_{\mathbf{kk}'}^{\rm SC}$
in the theory of the superconducting phase is analogous to the role of
$\Gamma_{\mathbf{kk}'}^{(1,2)}$
for the SDW, see
Sec.~\ref{sec::MF_SDW}.
Similar to
Eq.~(\ref{HintPsi})
we ignored screening retardation in
Eq.~(\ref{HintSCpsi})
as well. For more discussion, see
subsection~\ref{subsec::RPA_issues}.

We assume that our triplet ($p$-wave) superconducting state is
characterized by the following anomalous expectation values
\begin{equation}
\label{eq::SC_anomalous}
\eta^{\rm SC}_{\mathbf{k}}
=
\left\langle
	\gamma^{\phantom{\dag}}_{-\mathbf{k}3\uparrow}
	\gamma^{\phantom{\dag}}_{\mathbf{k}3\uparrow}
\right\rangle
=
\left\langle
	\gamma^{\phantom{\dag}}_{-\mathbf{k}3\downarrow}
	\gamma^{\phantom{\dag}}_{\mathbf{k}3\downarrow}
\right\rangle.
\end{equation}
This specific choice is but one possibility among many; others are
connected to
Eq.~(\ref{eq::SC_anomalous})
through unitary transformations
representing O(3) rotations of electron spin.
The superconducting order parameter can be defined as
\begin{equation}
\label{deltaSC}
\Delta^{\rm SC}_{\mathbf{k}}
=
\frac{1}{\cal N}\sum_{\mathbf{k}'}
	\Gamma_{\mathbf{kk}'}^{\rm SC} \eta^{\rm SC}_{\mathbf{k}'}\,.
\end{equation}
When momentum
$\mathbf{k}$
is close to the Dirac point
$\mathbf{K}_{\xi}$,
the expectation value
$\eta^{\rm SC}_{\mathbf{k}}$
couples electrons belonging to different valleys. Since
$\eta^{\rm SC}_{\mathbf{k}}$
couples electrons with the same spin, one has
\begin{eqnarray}
\label{eq::odd_orbital}
\eta^{\rm SC}_{-\mathbf{k}}=-\eta^{\rm SC}_{\mathbf{k}}
\quad
\Leftrightarrow
\quad
\Delta^{\rm SC}_{-\mathbf{k}}=-\Delta^{\rm SC}_{\mathbf{k}}.
\end{eqnarray}
Indeed, as the spin part of the Cooper pair wave function is even, the
orbital wave function must be odd.

Performing the standard mean-field decoupling in
Eq.~\eqref{HintSCpsi}
and minimizing the grand potential, we derive the zero-temperature
self-consistency equation for
$\Delta^{\rm SC}_{\mathbf{k}}$
\begin{equation}
\label{DeltaSCeq}
\Delta^{\rm SC}_{\mathbf{k}}
=
-\int\!\frac{d^2\mathbf{k}'}{v_{\rm BZ}}\,
	\frac{\Gamma_{\mathbf{kk}'}^{\rm SC} \Delta^{\rm SC}_{\mathbf{k}'}}
	{2\sqrt{
		\big[\varepsilon^{(3)}_{\mathbf{k}'}-\mu \big]^2
		+
		\big|\Delta^{\rm SC}_{\mathbf{k}'} \big|^2}}.
\end{equation}
The minus sign in the right-hand side of the self-consistency equation
is due to the repulsive Coulomb interaction. However, as we will show below the r.h.s. of Eq.~\eqref{DeltaSCeq} can be positive for the specific choice of the form of the order parameter.

The main contribution to the
integral in Eq.~\eqref{DeltaSCeq}
comes from the momenta $\mathbf{k}'$ near each Dirac point. In these regions
it is convenient to define
\begin{equation}
\tilde\varepsilon_{p}
=
\varepsilon^{(3)}_{\mathbf{K}_{\xi}+\mathbf{p}}\,,
\quad
\end{equation}
where
\begin{eqnarray}
\tilde\varepsilon_{p}\!
=\!
\sqrt{v_{\rm F}^2p^2\!+\!\frac{e^2\Phi^2}{4}\!+\!\frac{t_0^2}{2}\!-\!
\sqrt{\!v_{\rm F}^2p^2(e^2\Phi^2\!+\!t_0^2)\!+\!\frac{t_0^4}{4}}}
\quad
\end{eqnarray}
depends on the absolute value of the vector
$\mathbf{p}$. We propose the following ansatz for
$\Delta^{\rm SC}_{\mathbf{k}}$
\begin{equation}
\label{ansatzSC}
\Delta^{\rm SC}_{\mathbf{K}_{\xi}+\mathbf{p}}=\left\{\begin{array}{cl}
\tilde\Delta^{\rm SC}_{\xi p}\cos\phi_{\mathbf{p}}&,\;\;|\mathbf{p}|<K_0\\
0&,\;\;\text{otherwise}
\end{array}\right.\,.
\end{equation}
where
$\tilde\Delta^{\rm SC}_{\xi p}$
depends only on the absolute value of vector
$\mathbf{p}$.
We can show that near Dirac points the following relation is true
\begin{equation}
\label{gggg}
\Gamma_{\mathbf{K}_{\xi}+\mathbf{p}\mathbf{K}_{\xi}+\mathbf{p}'}^{\rm SC}
=
\Tilde \Gamma^{\rm SC} (p,\,p',\,\phi_{\mathbf{p}}-\phi_{\mathbf{p}'})
\end{equation}
In this formula, the function
$\Tilde \Gamma^{\rm SC} (p,\,p',\,\phi_{\mathbf{p}}-\phi_{\mathbf{p}'})$
depends on the absolute values of the vectors
$\mathbf{p}$
and
$\mathbf{p}'$,
and the polar angle
$\phi_{\mathbf{p}}-\phi_{\mathbf{p}'}$
between them. Then, it is easy to derive
\begin{equation}
\label{wwwwwww}
\int\limits_{0}^{2\pi}\frac{d\phi'}{2\pi}
	\Tilde \Gamma^{\rm SC} (p,\,p',\,\phi_{\mathbf{p}}-\phi')\cos\phi'
=
-\cos\phi_{\mathbf{p}}W(p,\,p')\,,
\end{equation}
where the kernel $W$ is
\begin{equation}
\label{W}
W(p,\,p')
=
-\int\limits_{0}^{2\pi}\frac{d\phi}{2\pi}
	\Tilde \Gamma^{\rm SC} (p,\,p',\,\phi)\cos\phi\,.
\end{equation}
The value
$W(p,\,p')$
depends on the interactions
$V^{ij}_{\mathbf{p}-\mathbf{p}'}$,
which in turn depend on
$q=|\mathbf{p}-\mathbf{p}'|$.
The most important for us here is that, when $q$ increases from
0 to
$\sim2k_{\rm F}^{(2)}$,
the functions
$V^{ij}_{\mathbf{q}}$
demonstrate a growing trend (see
Fig.~\ref{FigV}).
As a result, the integral
$\int d\phi\tilde{\Gamma}\cos\phi$
in
Eq.~\eqref{W}
is negative at sufficiently small $p$ and $p'$, making
$W(p,\,p')$
positive at small $p$, $p'$. Taking into account Eqs.~\eqref{ansatzSC}, \eqref{gggg}, \eqref{wwwwwww}, \eqref{W}, and neglecting the intervalley coupling,
we can rewrite the self-consistency equation~\eqref{DeltaSCeq} in the form
\begin{equation}
\label{DeltaXieq}
\tilde\Delta^{\rm SC}_{\xi p}
=
\frac{\pi}{v_{\rm BZ}}\int\limits_{0}^{K_0}p'dp'\,
	\frac{W(p,p')\tilde\Delta^{\rm SC}_{\xi p'}}
	{\sqrt{
		(\Tilde \varepsilon_{p'}-\mu)^2
		+
		(\tilde\Delta^{\rm SC}_{\xi p'})^2
		}
	}.
\end{equation}
\begin{figure}[t]
\centering
\includegraphics[width=0.95\columnwidth]{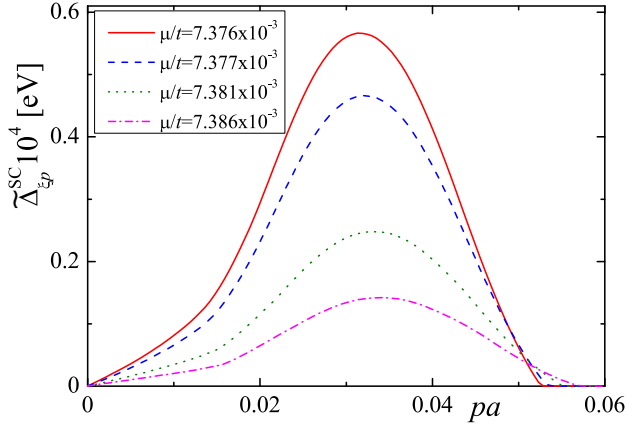}
\caption{
\label{FigDeltaK}
The dependencies of the superconducting order parameter
$\tilde\Delta^{\rm SC}_{\xi p}$
on $p$ calculated at
$e\Phi =0.1t_0=40$\,meV
and at four different values of the chemical potential (see the legend in
figure). For all curves we take $\epsilon=1$.
}
\end{figure}
We solve this integral equation numerically using successive iterations
technique. The typical curves
$\tilde\Delta^{\rm SC}_{\xi p}$
versus $p$, calculated for
$e\Phi =0.1t_0=40$\,meV,
$\epsilon=1$,
and several values of $\mu$, are plotted in
Fig.~\ref{FigDeltaK}.
In this figure we observe that, as $p$ grows, the function
$\tilde\Delta^{\rm SC}_{\xi p}$
first increases from zero, then, passing the maximum, and goes back to zero
when
$p\approx4k_{\rm F}^{(2)}$.
The order parameter
$\tilde\Delta^{\rm SC}_{\xi p}$
vanishes at
$p=0$
because the integral over $\phi$ in
Eq.~\eqref{W}
is zero when
$p=0$.
At momenta
$p\gtrsim4k_{\rm F}^{(2)}$,
we have
$\tilde\Delta^{\rm SC}_{\xi p}=0$
because the function
$W(p,\,p')$
is negative at sufficiently large $p$ and $p'$.

\begin{figure}[t]
\centering
\includegraphics[width=0.95\columnwidth]{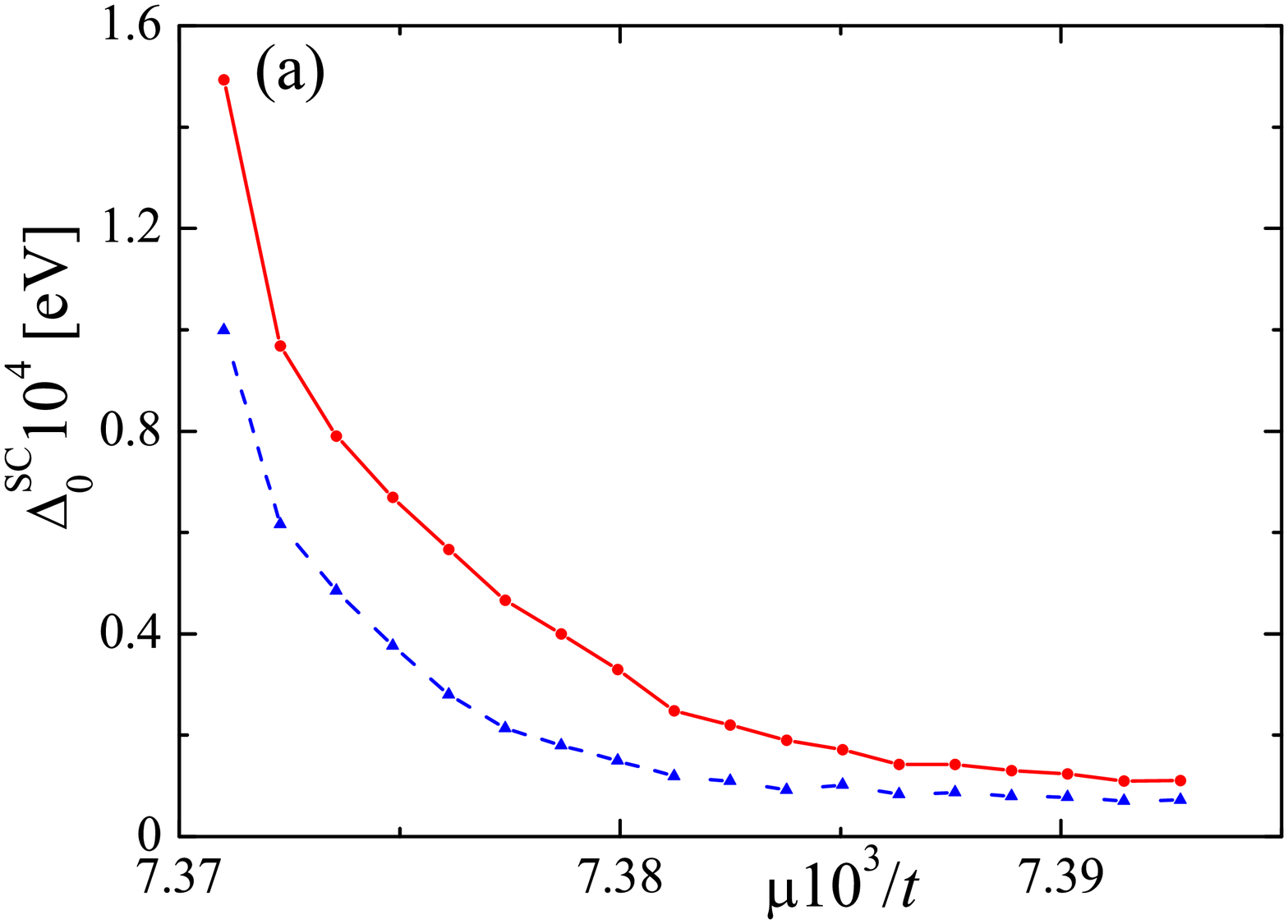}
\includegraphics[width=0.95\columnwidth]{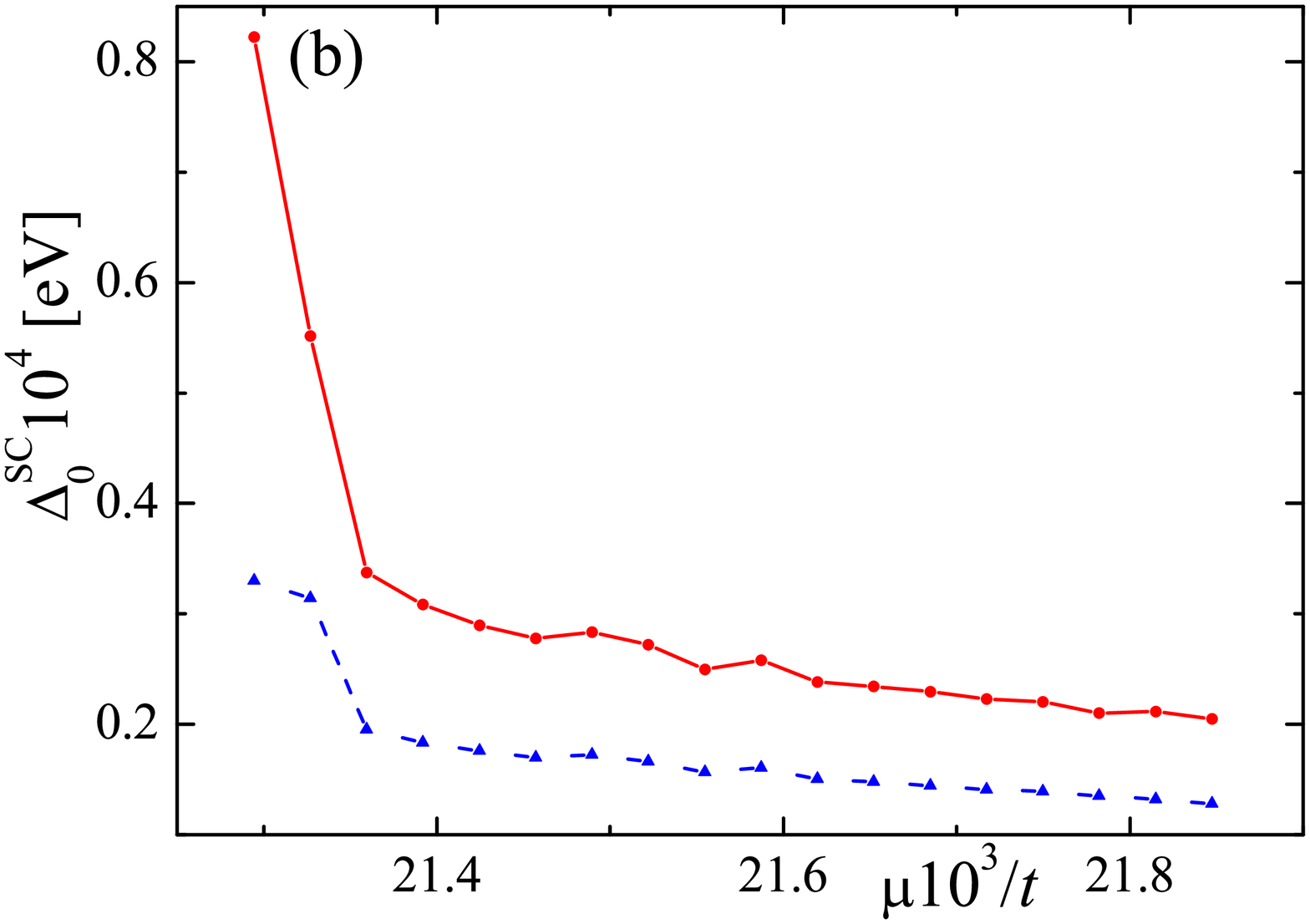}
\caption{
\label{FigMaxDelta}
The dependence of the
$\Delta^{\rm SC}_0\equiv\max(\tilde\Delta^{\rm SC}_{\xi p})$
on the chemical potential $\mu$ calculated for various values of the bias
voltage
$e\Phi$
and dielectric constant $\epsilon$.
Panel~(a) presents the data for
$e\Phi=0.1t_0=40$\,meV,
the curves in panel~(b) are plotted for
$e\Phi=0.3t_0=120$\,meV.
In both panels, the (red) solid curves with filled circles correspond to
$\epsilon=1$.
The (blue) dashed curves with filled triangles correspond to
$\epsilon=5$.
}
\end{figure}

Figure~\ref{FigMaxDelta}
shows the dependence of
$\Delta^{\rm SC}_0\equiv\max(\tilde\Delta^{\rm SC}_{\xi p})$
on the chemical potential. The value of
$\Delta^{\rm SC}_0$
decreases with the increase of the chemical potential. We attribute such a
behavior to the fact that the density of states at the Fermi level
decreases with $\mu$. Experimental
data~\cite{SCBLGNature2022}
also suggests that the large density of states is crucial for the
superconductivity. The data in
Fig.~\ref{FigMaxDelta}
indicate that, similar to the SDW case, the superconductivity weakens when
$\epsilon$ increases.

The numerical results shown in
Fig.~\ref{FigMaxDelta}
demonstrate that
$\Delta^{\rm SC}_0$
can be as large as several hundreds of mK, which exceeds by an order of magnitude the
superconducting transition temperature $T_c=26$\,mK
measured experimentally~\cite{SCBLGNature2022}.
To reconcile the theory with the experiment, let us
estimate $T_c$ for our model.
The finite-temperature generalization of the self-consistency equation~\eqref{DeltaXieq}
was derived using a standard technique and it differs
from the equation for $T=0$ only by multiplication of the function under integral by
$\tanh\left[\sqrt{(\varepsilon_{p'}-\mu)^2+\Delta_{\xi p'}^2}/(2T)\right]$.
We solve numerically the  self-consistency equation
for $\tilde\Delta^{\rm SC}_{\xi p}$ at finite temperature $T$ for several values of $\mu$ and
observe a significant disparity between the order parameter and the transition temperature.
For example, at $\mu/t=7.374\times10^{-3}$,
we could not find a non-trivial solution
$\tilde\Delta^{\rm SC}_{\xi p} \ne 0$
when $T>23$\,mK, which is $\approx 0.11 \Delta_0^{\rm SC} \ll \Delta_0^{\rm SC}$.
Therefore, for this value of $\mu$ the transition
temperature is about $23$\,mK, which agrees well with the experiment.
We associate that ``strange" feature of our model, $T_c \ll \Delta_0^{\rm SC}$,  with the fact that the considered Fermi sea in the AB-BLG is
very shallow: the Fermi energy defined as
$\varepsilon_{\rm F}=\mu-\mu_{\text{min}}$
is comparable or even smaller than
$\Delta^{\rm SC}_0$.

\begin{figure}[t]
\centering
\includegraphics[width=0.95\columnwidth]{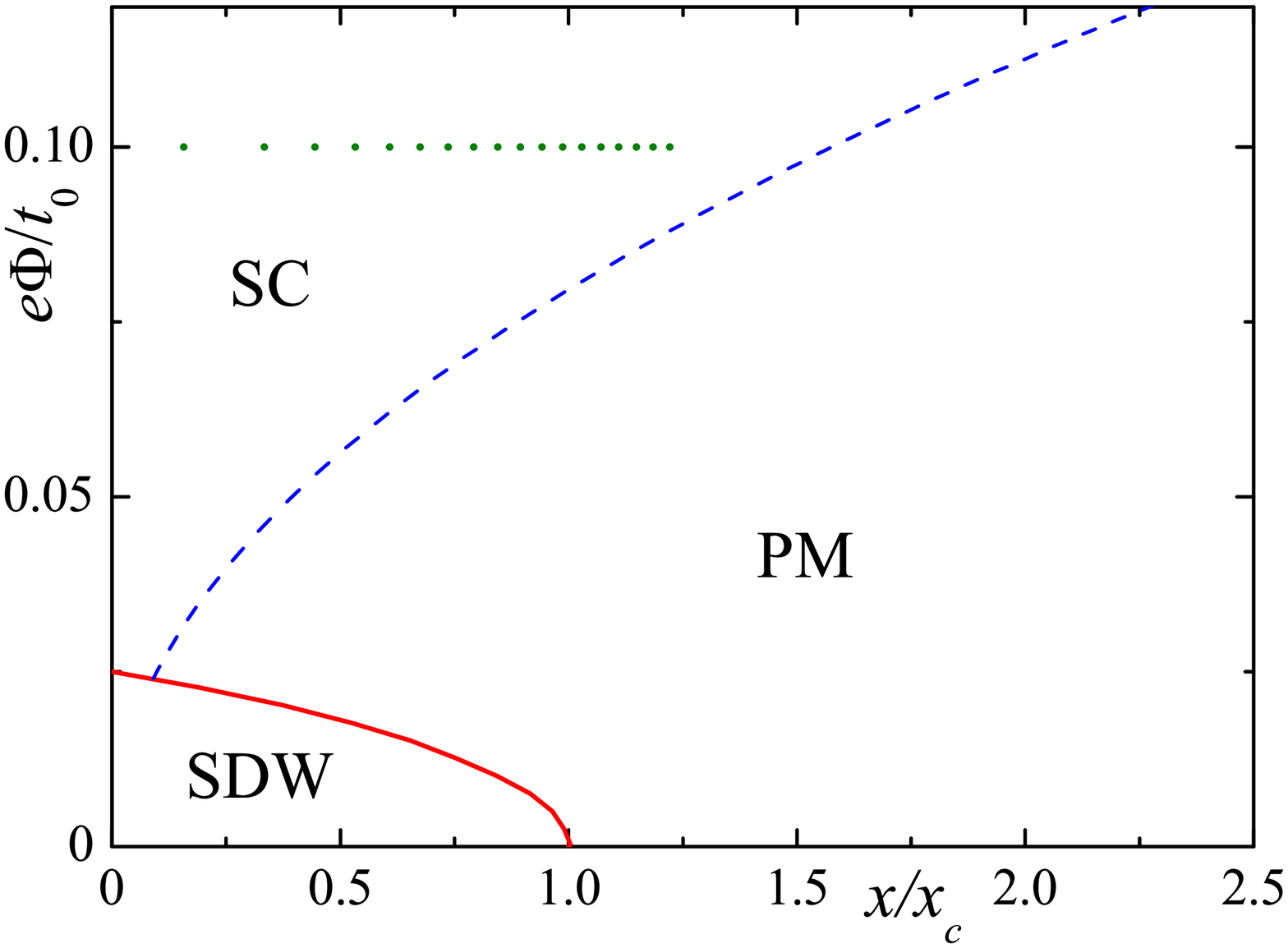}
\caption{
\label{FigPhDiag}
Schematic phase diagram of the system in the
$x$\,--\,$e\Phi$
plane. Red solid curve separates the SDW state from SC and PM states. Blue
dashed curve is the curve of the crossover between the SC and PM states.
The critical doping $x_c$ corresponding SDW to PM transition at
$e\Phi=0$
is estimated as
$x_c\approx1.3\times10^{-5}$
(for $t=2.7$\,eV, $t_0=0.4$\,eV,
and
$\epsilon=1$).
Green circles depict the points ($x,\,e\Phi$) at which
the superconducting order parameter was calculated numerically [see
Fig.~\ref{FigMaxDelta}(a)].
}
\end{figure}

All the results above can be summarized in the phase diagram of the model
in the
$x$\,--\,$e\Phi$
plane. Let us consider first the SDW phase. For a given bias voltage
$e\Phi$ the critical chemical potential
$\mu_c^{\text{SDW}}$,
above which the SDW state is suppressed, can be found from the equation
[compare it with
Eq.~\eqref{SDWcriterion}]
\begin{equation}
\label{dopedSDWcriterion}
\int\limits_{|\mathbf{p}|<K_0}\!\!\!\!
	\frac{d^2\mathbf{p}}{v_{\rm BZ}}\,
	\frac{
		\Tilde \Gamma^{(1)}_{0\mathbf{p}}
		+
		\Tilde \Gamma^{(2)}_{0\mathbf{p}}
	}{2\Tilde\varepsilon^{(3)}_{\mathbf{p}}}
	\Theta\left(\Tilde\varepsilon^{(3)}_{\mathbf{p}}-\mu\right)=1\,.
\end{equation}
Solving this equation and using
Eq.~\eqref{doping}
we obtain the curve
$x_c^{\text{SDW}}
=
x_c^{\text{SDW}}(e\Phi)$
separating the SDW phase from superconducting (SC) and paramagnetic (PM)
phases. In connection with the SC state, we restrict ourselves by
considering the chemical potentials
$\mu<e\Phi/2$
when the system has two Fermi surface sheets near each Dirac point. The
case of larger chemical potentials requires separate analysis. Note,
however, that for
$\mu>e\Phi/2$
the superconducting order parameter, even if non-zero, will be small [see
Eq.~\eqref{vFs}
and the text below it]. Thus, one can consider the curve
$x_c^{\text{SC}}(e\Phi)=x(e\Phi/2)\approx e^2\Phi^2/(2\pi\sqrt{3}t^2)$
as the curve of the crossover between SC and PM states. The resultant phase
diagram is shown in
Fig.~\ref{FigPhDiag}.

\section{Discussion and conclusions}
\label{sec::discussion}

We argued above that the doped and biased Bernal stacked bilayer graphene
can host Coulomb-interaction-driven triplet superconducting state. In this
section we will discuss certain important details of the mechanism that
remain untouched in the more formal presentation.

\subsection{Kohn-Luttinger roots of the superconductivity}

The superconducting state becomes stable thanks to the fact that the
functions
$V^{ij}_{\mathbf{q}}$
increase with $q$ at small transferred momenta. Such a behavior of
$V^{ij}_{\mathbf{q}}$
is obtained with the help of RPA.
As the RPA is an uncontrollable approximation, one may wonder if our
superconducting phase is indeed a genuine article, and not an artifact of
careless theoretical assumptions. To such concerns we offer twofold
redress. For one, the RPA validity is discussed below, see
subsection~\ref{subsec::RPA_issues}.

Beside this, we argue that our mechanism of the superconductivity is not
rooted in particulars of the RPA approach. Rather, one can trace its
origins to the
proposal~\cite{kohn_lutt}
of Kohn and Luttinger (KL). It is instructive to compare the two
mechanisms. Unlike our RPA-based formalism, the classical KL
calculations~\cite{kohn_lutt}
rely on the second-order perturbation theory in powers of the bare Coulomb
interaction. Since the second-order correction represents screening, it
reduces the electron-electron repulsion. Loosely speaking, it is a kind of
attraction that counteracts the bare Coulomb repulsion. Further, this
correction is singular due to the Kohn anomaly in the polarization
operator. The KL paper demonstrated that, for sufficiently large Cooper
pair orbital momentum, the polarization operator, being singular, overcomes
the non-singular bare Coulomb interaction. In such an orbital channel
effective attraction emerges, leading to the superconducting instability.

The second-order correction, as a separate theoretical object, does not
occur in our formalism. However, similar to the KL idea, the role of the
polarization operator is quite essential for our mechanism as well. We see
that the strong screening at low $q$ dominates in the effective interaction, as
attested by the curves in
Fig.~\ref{FigV}.
(This is particularly true for
$V^{11}$ and $V^{12}$.)
The polarization operator, controlling the renormalized interaction at
small $q$, causes the overall growth of the effective interactions for
growing $q$ in the interval
$0.025 < qa < 0.1$.
The latter growth of
$V^{ij}_{\bf q}$
is the cornerstone of the mechanism suggested in
Sec.~\ref{sec::superconducting_mech}.

\subsection{RPA validity}
\label{subsec::RPA_issues}

Let us briefly discuss to which extent the static RPA interaction can be
considered as a reliable approach for our purposes. This problem contains two
sub-problems: (i) Is the RPA by itself is reliable in our situation? (ii)
Is the static version of the RPA effective interaction can be used to study
the SDW and superconductivity?

In connection to~(i) let us consider the following. It is generally
accepted that the RPA works well for phenomena involving distances greater
than a characteristic screening (Debye) length 
$l_D$~\cite{PhysRev.85.338,mahan2000many,bruus2004many}.
From the data shown in 
Fig.~\ref{FigV}, 
we can conclude that $l_D$ is of the order of $10\,a$, while the
superconducting and SDW orders are determined mostly by the structure of
the screened Coulomb interaction on the scales larger than
$l_D$.
From our numerical results it follows that the superconducting coherence
length
$\xi^{\textrm{SC}}\sim\hbar v_{\textrm{F}}/\Delta^{\textrm{SC}}$
is about $100\,a$ in the parameters range of interest, which is larger than
$l_D$.
Moreover, it is commonly believed that using the RPA approach is especially
reasonable for the graphene-based systems since each bubble diagram enters
the RPA expansion with a degeneracy factor $N_d=4$ (this is due to the spin
and valley
degeneracies)~\cite{InteractionsInGrapheneRevModPhys2012}.

(ii)~The use of static effective interaction, as expressed in
Eqs.~(\ref{HintPsi})
and~(\ref{HintSCpsi}),
is valid as long as the full dynamic polarization operator
$\Pi_{\bf q}^{ij} (\omega)$
does not vary significantly over the frequency scale set by the order
parameter. For the SDW phase, the order parameter is several meV. Does
$\Pi_{\bf q}^{ij} (\omega)$
for the undoped AB-BLG varies strongly over this scale? To answer this
question, we want to make a simple observation. The only parameters
entering
$\Pi_{\bf q}^{ij} (\omega)$
are $t$ and
$t_0$,
both of which are much larger than
$\Delta^{\rm SDW}$.
This indicates clearly that, for $\omega$ limited to the interval whose
width is of order of
$\Delta^{\rm SDW}$,
the dynamical polarization operator may be safely approximated by its
static version.

The situation with the superconducting phase requires more diligence: since
the superconductivity is observed under the doping, in addition to the
tunneling amplitudes, the Fermi energy enters
$\Pi_{\bf q}^{ij} (\omega)$.
Since
$\varepsilon_{\rm F}$
is the smallest of the three energy parameters in
$\Pi_{\bf q}^{ij} (\omega)$,
we conclude that, when the superconducting energy scale does not exceed
$\varepsilon_{\rm F}$,
the static approximation works well.

\subsection{Magnetic field effect}

In experiment~\cite{SCBLGNature2022}
a superconducting state was observed only at finite in-plane
magnetic field. This finding supports our assumption about the triplet
structure of the superconducting order parameter. Indeed, it is known that
the $p$-wave superconducting state, unlike its singlet counterpart,
possesses a finite paramagnetic (Zeeman)
susceptibility~\cite{mineev1999introduction}.
Consequently, the $p$-wave superconductivity is much more robust against
applied magnetic field. We can speculate that, in the experiment, at finite
applied field, the superconducting state replaces a non-superconducting
phase that has lower zero-field energy but weaker Zeeman susceptibility.
In such a scenario, application of the field can invert relative stability
of the two phases, leading to the realization of the superconductivity,
which is metastable at zero field.

The nature of the phase supplanted by the superconductivity is an
interesting question worth further research. For example, this phase can be
one of several fractional metallic states (doped SDW with spin- and
valley-polarized Fermi surface), considered theoretically for graphene
bilayer systems in
Refs.~\onlinecite{rakhmanov2023QM,sbyocha2021FRAM}.
Experimental results in 
Ref.~\onlinecite{SCBLGNature2022}
support such a hypothesis.

\subsection{Other types of superconducting order parameter}

The superconducting order parameter discussed above is not the only
possible, as other types of superconductivity might be stabilized in our
AB-BLG model. To illustrate this point, consider the following reasoning.
The suggested above anomalous expectation
$\eta^{\rm SC}_{\mathbf{k}}
=
\langle\gamma^{\phantom{\dag}}_{-\mathbf{k}3\sigma}
\gamma^{\phantom{\dag}}_{\mathbf{k}3\sigma}\rangle$
couples electrons in different valleys and the total momentum of the Cooper
pair is zero. One can consider another choice, when both electrons
constituting a pair belong to the same valley. The corresponding
expectation value is
$\Tilde \eta^{\rm SC}_{\mathbf{p} \xi}
=
\langle
	\gamma^{\phantom{\dag}}_{\mathbf{K}_{\xi}-\mathbf{p}3\sigma}
	\gamma^{\phantom{\dag}}_{\mathbf{K}_{\xi}+\mathbf{p}3\sigma}
\rangle$.
The total momentum of such a pair is
$2\mathbf{K}_{\xi}$.
Consequently, the superconducting order parameter oscillates in real space,
making this state a type of pair-density
wave~\cite{pdw2004theory}.

Since we limit ourselves to small doping, only one band crosses the Fermi
level. The situation becomes richer at stronger doping, when two bands are
partially filled. When this happens, an inter-band order parameter may be
defined. It also oscillates in real space. However, absence of van Hove
singularities at higher $\mu$ implies that the corresponding condensation
energy is low.

In general, the valley degeneracy is a peculiar feature of the
graphene-based materials, which introduces additional complications in the
task of superconducting phases classification.
Recent work~\cite{su4_classification}
on the classification of non-superconducting phases in graphene bilayer
demonstrated the challenges that one faces when the discrete index space
grows twofold (from twofold spin degeneracy of a BCS-like models to
fourfold spin-valley degeneracy of graphene-based metals). In
Ref.~\onlinecite{su4_classification}
we offer an SU(4)-based approach to the non-superconducting-order
classification that could possibly be extended to the superconducting
phases, as well.

\subsection{Trigonal warping}

The single-particle Hamiltonian of our model is constructed under the
assumption that the inter-layer hopping occurs only between
nearest-neighbor inter-layer sites located in positions $1A$ and $2B$, and
more distant inter-layer hoppings are neglected. When this simplification is
lifted, the low-energy electronic spectrum experiences certain
modifications. For example, in the case 
$e\Phi=0$, 
if we include the hopping amplitude 
$t_3$ 
between nearest-neighbor sites in positions $1B$ and $2A$ (see, e.g.,
Ref.~\onlinecite{bilayer_review2016}), 
two parabolic bands touching each other at the Dirac points are converted
to four Dirac cones located near the Dirac points. Such a low-energy
structure is called trigonal warping. Incorporation of the trigonal warping
in our model alters the results in some aspects. First, it can change the
estimate of the value of the SDW order parameter. Strictly speaking,
Eq.~\eqref{DeltaSDWtrans}
has a solution for arbitrary small interaction, since the integral in
right-hand side of this equation diverges logarithmically when
$\Delta^{\text{SDW}}_{\xi}\to0$.
If the trigonal warping is taken into account, the non-trivial solution to
Eq.~\eqref{DeltaSDWtrans}
appears only at finite interaction strength, since the density of states
vanishes at zero energy. However, the analysis reveals that the interaction
is rather strong, while the trigonal warping modify the electron spectrum
only at energies about $1$\,meV (see, e.g.,
Ref.~\onlinecite{bilayer_review2016}),
thus, we expect that the estimate for the SDW order parameter does not
change substantially when the trigonal warping is accounted for.

The trigonal warping, of course, transforms the low-energy spectrum of the
AB-BLG, which affects the superconducting state. We believe, however, that
the trigonal warping does not change our results qualitatively. Studies of
the superconducting state via the renormalized Coulomb interaction, which
take into account the trigonal warping, have been reported in
Refs.~\onlinecite{Guinea2023superconductivity,wagner2023superconductivity}.
The characteristic critical temperatures found there are consistent with
our results.

\subsection{The role of the order parameter fluctuations}

In two-dimensional systems, finite-temperature fluctuations of the
Goldstone modes destroy any non-Ising long-range order. Specifically, in
the SDW
phase~\cite{aa_graph_pha_sep_sboycha2013}
the Goldstone mode is the spin-wave excitations described by the O(3)
non-linear $\sigma$ model in 2+1-dimensional space. As temperature grows,
the O(3) field correlations smoothly decay. As a result, a continuous
transition, expected within the mean-field framework, is replaced by a
smooth crossover. It is expected that the characteristic crossover
temperature
$T_*^{\rm SDW}$
is of order of the mean field transition temperature
\begin{eqnarray}
\label{eq::SDW_xover}
T_*^{\rm SDW} \sim T_{\rm MF}^{\rm SDW} \sim \Delta^{\rm SDW}.
\end{eqnarray}
This relation indicates that, for
$0 < T \ll \Delta^{\rm SDW}$
robust signatures of short-range SDW order must be detectable
experimentally.

Now we discuss the superconducting state. Since the bilayer sample is very
thin, the magnetic field screening by the superconducting currents may be
neglected. In such a situation, the fluctuations of the complex phase of
$\Delta^{\rm SC}$
can be described by the XY non-linear $\sigma$ model. At sufficiently large
$T$ this model demonstrates the Berezinskii-Kosterlitz-Thouless transition
whose critical temperature, similar to
estimate~(\ref{eq::SDW_xover}),
is of the order of the mean field critical temperature
$T_c$.

\subsection{The role of substrate's dielectric constant}

Our calculations show that both the SDW and the superconductivity are
weakened when the dielectric constant of the substrate grows. This is a
straightforward consequence of the fact that both ordered states rely on
the long-range Coulomb interaction. At the same time, these phases have
dissimilar sensitivities to the increase of $\epsilon$. Specifically, we
have seen that the growth of the dielectric constant from
$\epsilon = 1$
to
$\epsilon = 5$
suppresses the SDW order parameter by more than an order of magnitude.
In contrast, the characteristic values of the superconducting order
parameter decrease roughly twofold at most. This suggests that a substrate
with larger dielectric constant shifts the balance between the SDW and the
superconductivity in favor of the latter. Such a possibility can be tested
experimentally. One must remember, however, that here we ignore short-range
interactions, which are insensitive to screening but may affect the
properties of the ordered states. If these contributions are indeed
significant in AB-BLG, the expected effect of $\epsilon$ might be weak.

\subsection{Conclusions}

In this paper we suggested a mechanism of superconductivity in the AB-BLG.
This mechanism is based on the renormalized Coulomb electron-electron
repulsion, and is similar in certain aspects to the Kohn-Luttinger
mechanism. The superconducting state competes against the spin-density wave
state, which is also stabilized by the Coulomb interaction. The
superconductivity in the proposed model has a $p$-wave structure. Our
estimate for the critical temperature, as well as order parameter
sensitivity to the doping, is consistent with recent experiment. Likewise,
in-plane magnetic field as a stabilization factor of the superconducting
phase fits to the proposed theoretical framework.

\section*{Acknowledgments}
This work is supported by RSF grant No.~22-22-00464,
\url{https://rscf.ru/en/project/22-22-00464/}.

\appendix

\section{Charge-conjugation symmetry of biased undoped AB-BLG}
\label{append::charge_conj}
%
\begin{figure}[t]
\centering
\includegraphics[width=0.95\columnwidth]{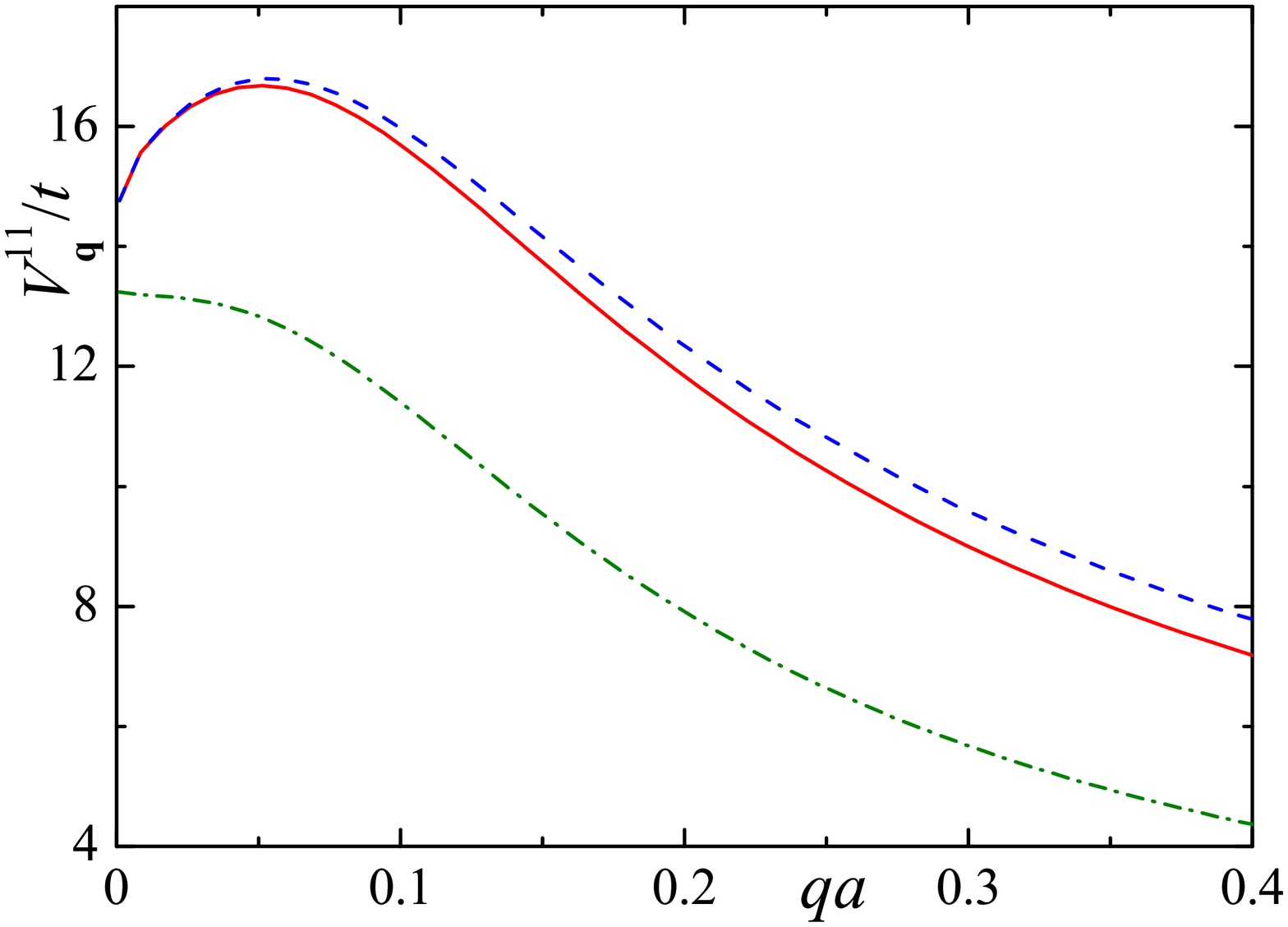}
\includegraphics[width=0.95\columnwidth]{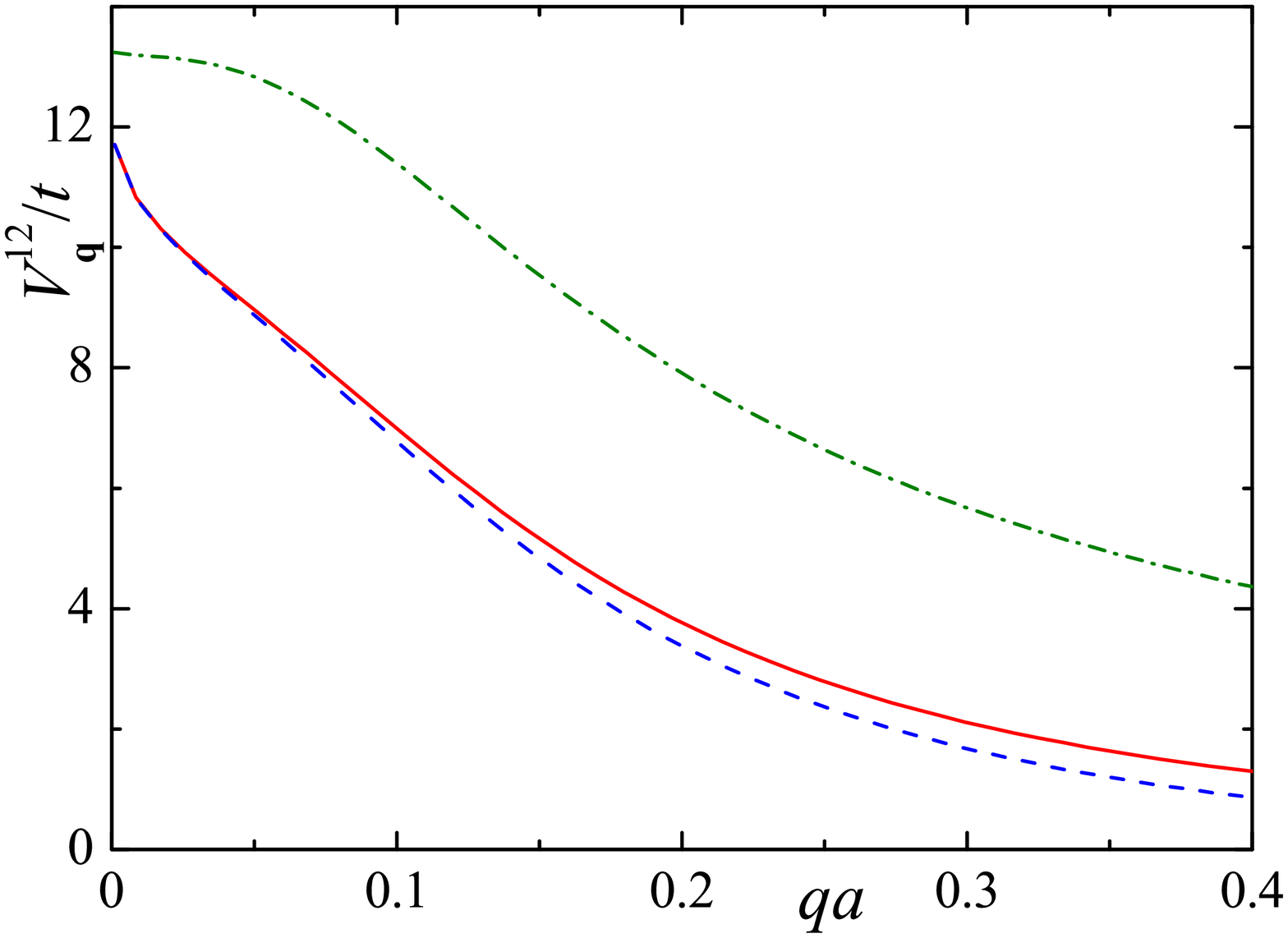}
\caption{The renormalized interaction components $V_{\mathbf{q}}^{ij}$ as functions of $q$,
calculated using different approximations; $\mathbf{q}=q(1,\,0)$, $\mu=0$, and $e\Phi=0$.
Panel~(a) shows $V^{11}_{\bf q}$, while panel~(b) shows $V^{12}_{\bf q}$.
In both panels the RPA interaction components,
Eqs.~(\ref{V11})
and~(\ref{V12}),
are plotted as (red) solid curve, approximate
expressions~(\ref{eq::append_V11})
and~(\ref{eq::append_V12})
are represented by (blue) dashed curves. Dash-dotted (green) curves
correspond to
approximation~(\ref{eq::append_V_ij_simple}).
We see that
Eqs.~(\ref{eq::append_V11})
and~(\ref{eq::append_V12})
work quite well for small $q$, while
formula~(\ref{eq::append_V_ij_simple})
is very crude approximation for our parameters choice.
\label{fig::append_approximat_formulas}
}
\end{figure}

Here we prove that our model is invariant under a certain
charge-conjugation transformation. This invariance explains why the
polarization operator components
$\Pi^{11}_{\mathbf{q}}$
and
$\Pi^{22}_{\mathbf{q}}$
are equal to each other as long as AB-BLG remains undoped. To start the
discussion we re-write the matrix
${\cal H}_{\bf k}$
from
Eq.~(\ref{Hkin})
as follows
\begin{eqnarray}
&&{\cal H}_{\bf k}
=
\frac{e \Phi}{2} \tau_z - t {\bf f}\cdot \boldsymbol{\nu}
\\
\nonumber
&&+ \frac{t_0}{4}
\left[
	(\nu_x + i\nu_y) (\tau_x + i \tau_y)
	+
	(\nu_x - i\nu_y) (\tau_x - i \tau_y)
\right],
\end{eqnarray}
where the Pauli matrices
$\nu_i$
act in the sublattice space, while another set of Pauli matrices
$\tau_i$
acts in the layer space, and
${\bf f}\cdot \boldsymbol{\nu} = \nu_x {\rm Re}\, (f_{\bf k} )
- \nu_y {\rm Im}\, (f_{\bf k})$.
For such a matrix an equality
\begin{eqnarray}
\label{eq::append_conjugation}
\nu_y \tau_x {\cal H}_{\bf k}^* \nu_y \tau_x
=
- {\cal H}_{\bf k}
\end{eqnarray}
holds true. This relation is the signature of the charge conjugation
symmetry. To reveal the invariance of
$H_0$
under the charge conjugation, we turn our attention to the
second-quantization formalism. We write
\begin{eqnarray}
\label{eq::appendix_H0}
H_0 = \sum_{{\bf k} \sigma} \sum_{\zeta\zeta'}
	\left[{\cal H}_{\bf k}\right]_{\zeta \zeta'}
	d^\dag_{{\bf k} \zeta \sigma }
	d^{\vphantom{\dag}}_{{\bf k} \zeta' \sigma },
\end{eqnarray}
where
$\left[{\cal H}_{\bf k}\right]_{\zeta\zeta'}$
are matrix elements of
${\cal H}_{\bf k}$,
and summation variables $\zeta$, $\zeta'$ are multi-indices containing
layer and sublattice labels:
$\zeta = (i, \alpha)$.

If we apply a charge conjugation Bogolyubov transformation
\begin{eqnarray}
\label{eq::append_Bogolyubov}
d^{\vphantom{\dagger}}_{{\bf k} \zeta \sigma}
\leftrightarrow
d^\dag_{{\bf k} \zeta \sigma},
\end{eqnarray}
the Hamiltonian
$H_0$
transforms to
\begin{eqnarray}
\label{eq::appendix_H1}
H^{\rm C}_0 = \sum_{{\bf k} \sigma} \sum_{\zeta\zeta'}
	\left[{\cal H}_{\bf k}\right]_{\zeta \zeta'}
	d^{\vphantom{\dag}}_{{\bf k} \zeta \sigma }
	d^\dag_{{\bf k} \zeta' \sigma }
\\
\nonumber
=
\sum_{{\bf k} \sigma} {\rm Tr}\, {\cal H}_{\bf k}
-
\sum_{{\bf k} \sigma} \sum_{\zeta\zeta'}
	\left[{\cal H}_{\bf k}\right]_{\zeta \zeta'}
	d^\dag_{{\bf k} \zeta' \sigma }
	d^{\vphantom{\dag}}_{{\bf k} \zeta \sigma }
\\
\nonumber
=
-\sum_{{\bf k} \sigma} \sum_{\zeta\zeta'}
	\left[{\cal H}_{\bf k}^*\right]_{\zeta\zeta'}
	d^\dag_{{\bf k} \zeta \sigma }
	d^{\vphantom{\dag}}_{{\bf k} \zeta' \sigma },
\end{eqnarray}
where we used the fact that
${\cal H}_{\bf k}$
has zero trace for any
${\bf k}$,
and
$\left[{\cal H}_{\bf k}\right]_{\zeta\zeta'}
=
\left[{\cal H}_{\bf k}^*\right]_{\zeta'\zeta}$
due to hermiticity. Thus
\begin{eqnarray}
H^{\rm C}_0
&=&
- \sum_{\mathbf{k}\sigma}
	\psi^{\dag}_{\mathbf{k}\sigma}
	{\cal H}_{\mathbf{k}}^*
	\psi^{\phantom{\dag}}_{\mathbf{k}\sigma}.
\end{eqnarray}
Defining new operator vector
$\psi^{\rm C}_{\mathbf{k}\sigma}$
by the relation
\begin{eqnarray}
\label{eq::append_new_operator}
\psi^{\phantom{\dag}}_{\mathbf{k}\sigma}
=
\nu_y \tau_x \psi^{\rm C}_{\mathbf{k}\sigma}
\end{eqnarray}
we can confirm, using
Eq.~(\ref{eq::append_conjugation}),
that
$H^{\rm C}_0$
is unitary equivalent to
$H_0$.

At the same time, in the first-quantization formalism,
Eq.~(\ref{eq::append_conjugation})
implies that the transformation
$\nu_y \tau_x \Psi^{(S)*}_{\bf k}$
converts the bispinor eigenvector
$\Psi^{(S)}_{\bf k}$
corresponding to the energy 
$\varepsilon^{(S)}_{\bf k}$
into another bispinor eigenvector representing
$-\varepsilon^{(S)}_{\bf k}$.
Examining our
definitions~(\ref{spec})
one can check that
$-\varepsilon^{(S)}_{\bf k}
=
\varepsilon^{(5-S)}_{\bf k}$.
Thus, it is convenient to introduce the abbreviation
$\bar{S} = 5 - S$.
It allows us to write
$\varepsilon^{(\bar{S})}_{\bf k} = -\varepsilon^{(S)}_{\bf k}$
and
\begin{eqnarray}
\Psi^{(\bar{S})}_{{\bf k} 1 A}
=
-i \Psi^{(S)*}_{{\bf k} 2 B},
\quad
\Psi^{(\bar{S})}_{{\bf k} 1 B}
=
i \Psi^{(S)*}_{{\bf k} 2 A}.
\end{eqnarray} 
Substituting these formulas into 
expression~(\ref{P})
for
$\Pi^{11}_{\bf q}$,
and exploiting the relation
\begin{eqnarray} 
n_{\rm F} (\varepsilon)
-
n_{\rm F} (\varepsilon')
=
-\left[ 
	n_{\rm F} (-\varepsilon)
	-
	n_{\rm F} (-\varepsilon')
\right],
\end{eqnarray}
one can explicitly demonstrate that
$\Pi^{11}_{\bf q} = \Pi^{22}_{\bf q}$.

Note that the Hamiltonian of the doped system does not possess this
symmetry. Indeed, the charge conjugation inverts the sign of $\mu$,
making the whole Hamiltonian non-invariant.

\section{Approximate expression for the screened interaction}
\label{append::V_ij}

Let us investigate here the accuracy of
approximation~(\ref{eq::V_ij_simple}).
We assume that
$\Pi^{11}_{\bf q} = \Pi^{22}_{\bf q}$
due to the charge-conjugation symmetry. In the limit
$qd \ll 1$
we expand
$\exp(-qd) \approx 1 - qd$
to derive
\begin{widetext}
\begin{eqnarray}
V^{11}_{\mathbf{q}}
=
V^{22}_{\mathbf{q}}
=
\frac{\frac{1}{2} (1-{\cal E}\Pi^{22}_{\mathbf{q}}) {\cal E}}
{
	(qd) (1 + {\cal E} \Pi^{12}_{\bf q})
	-
	{\cal E}
	(
		1
		+
		\frac{1}{2}{\cal E} \Pi^{12}_{\bf q}
		-
		\frac{1}{2}{\cal E} \Pi^{11}_{\bf q}
	)
	(\Pi^{11}_{\bf q} + \Pi^{12}_{\bf q})
},
\label{eq::append_V11}
\\
V^{12}_{\mathbf{q}}
=
V^{21}_{\mathbf{q}}
=
\frac{\frac{1}{2} (1 - qd +{\cal E}\Pi^{12}_{\mathbf{q}}) {\cal E}}
{
	(qd) (1 + {\cal E} \Pi^{12}_{\bf q})
	-
	{\cal E}
	(
		1
		+
		\frac{1}{2}{\cal E} \Pi^{12}_{\bf q}
		-
		\frac{1}{2}{\cal E} \Pi^{11}_{\bf q}
	)
	(\Pi^{11}_{\bf q} + \Pi^{12}_{\bf q})
},
\label{eq::append_V12}
\end{eqnarray}
\end{widetext}
where we introduce the energy scale
${\cal E} = 2 A d$.
For permittivity
$\epsilon = 5$
one finds
${\cal E} = 23.2$\,eV,
or, equivalently,
${\cal E} = 8.6t$.
This energy can be used to re-write
formula~(\ref{eq::V_ij_simple})
\begin{eqnarray}
\label{eq::append_V_ij_simple}
V^{ij}_{\mathbf{q}}
=
\frac{\frac{1}{2} {\cal E}}
	{(qd) - {\cal E} (\Pi^{11}_{\mathbf{q}}+\Pi^{12}_{\mathbf{q}})}.
\end{eqnarray}
Comparing this relation and
Eqs.~(\ref{eq::append_V11})
and~(\ref{eq::append_V12}),
one concludes that
Eq.~(\ref{eq::append_V_ij_simple})
is valid only when
${\cal E} \Pi^{ij}_{\bf q}$
is much smaller than unity at small $qd$. However, for our model
parameters, the quantity
${\cal E} \Pi^{ij}_{\bf q}$
is of order of unity, making
Eqs.~(\ref{eq::append_V_ij_simple}) and~(\ref{eq::V_ij_simple})
a poor approximation.
Figure~\ref{fig::append_approximat_formulas}
allows one to compare the accuracy of the two approximations.


\end{document}